\documentstyle[12pt]{article}
\begin{document}

\def\ds{\displaystyle}
\def\beq{\begin{equation}}
\def\eeq{\end{equation}}
\def\bea{\begin{eqnarray}}
\def\eea{\end{eqnarray}}
\def\beeq{\begin{eqnarray}}
\def\eeeq{\end{eqnarray}}
\def\ve{\vert}
\def\vel{\left|}
\def\ver{\right|}
\def\nnb{\nonumber}
\def\ga{\left(}
\def\dr{\right)}
\def\aga{\left\{}
\def\adr{\right\}}
\def\lla{\left<}
\def\rra{\right>}
\def\rar{\rightarrow}
\def\nnb{\nonumber}
\def\la{\langle}
\def\ra{\rangle}
\def\ba{\begin{array}}
\def\ea{\end{array}}
\def\tr{\mbox{Tr}}
\def\ssp{{\Sigma^{*+}}}
\def\sso{{\Sigma^{*0}}}
\def\ssm{{\Sigma^{*-}}}
\def\xis0{{\Xi^{*0}}}
\def\xism{{\Xi^{*-}}}
\def\qs{\la \bar s s \ra}
\def\qu{\la \bar u u \ra}
\def\qd{\la \bar d d \ra}
\def\qq{\la \bar q q \ra}
\def\gGgG{\la g^2 G^2 \ra}
\def\q{\gamma_5 \not\!q}
\def\x{\gamma_5 \not\!x}
\def\g5{\gamma_5}
\def\sb{S_Q^{cf}}
\def\sd{S_d^{be}}
\def\su{S_u^{ad}}
\def\ss{S_s^{??}}
\def\sbp{{S}_Q^{'cf}}
\def\sdp{{S}_d^{'be}}
\def\sup{{S}_u^{'ad}}
\def\ssp{{S}_s^{'??}}
\def\sig{\sigma_{\mu \nu} \gamma_5 p^\mu q^\nu}
\def\fo{f_0(\frac{s_0}{M^2})}
\def\ffi{f_1(\frac{s_0}{M^2})}
\def\fii{f_2(\frac{s_0}{M^2})}
\def\O{{\cal O}}
\def\sl{{\Sigma^0 \Lambda}}
\def\es{\!\!\! &=& \!\!\!}
\def\ap{\!\!\! &\approx& \!\!\!}
\def\ar{&+& \!\!\!}
\def\ek{&-& \!\!\!}
\def\cp{&\times& \!\!\!}
\def\se{\!\!\! &\simeq& \!\!\!}
\def\kpm{&\pm& \!\!\!}
\def\kmp{&\mp& \!\!\!}


\def\simlt{\stackrel{<}{{}_\sim}}
\def\simgt{\stackrel{>}{{}_\sim}}


\title{
         {\Large
                 {\bf
          General analysis of CP violation in polarized
                  $b \rar d \ell^+ \ell^-$ decay 
                 }
         }
      }

\author{\vspace{1cm}\\
{\small T. M. Aliev \thanks
{e-mail: taliev@metu.edu.tr}\,\,,
V. Bashiry
\,\,,
M. Savc{\i} \thanks
{e-mail: savci@metu.edu.tr}} \\
{\small Physics Department, Middle East Technical University,
06531 Ankara, Turkey} }

\date{}

\begin{titlepage}
\maketitle
\thispagestyle{empty}

\begin{abstract}
The CP violating asymmetries in the $b \rar d \ell^+ \ell^-$ decay, when one
of the leptons is polarized, is investigated using the most general form of
the effective Hamiltonian. The sensitivity of the CP violating asymmetries
on the new Wilson coefficients is studied.
\end{abstract}

~~~PACS numbers: 12.60.--i, 14.65.Fy 
\end{titlepage}

\section{Introduction}

Rare B meson decays, induced by the flavor changing neutral current (FCNC)
$b \rar s(d)$ transitions, provide one the most promising research area in
particle physics. Interest to rare B meson decays has its roots in their
role being potentially the precision testing ground 
for the standard model (SM) at loop level and looking for new physics beyond
the SM \cite{R5701}. Experimentally these decays will provide a more precise
determination of the elements of the Cabibbo--Kobayashi--Maskawa (CKM)
matrix, such as, $V_{tq}~(q=d,s,b)$, $V_{ub}$ and CP violation. In FCNC
decays, any deviation over the SM results is an unambiguous indication for
new physics. The first observation of the radiative $B \rar X_s \gamma$
decay by CLEO \cite{R5702}, and later by ALEPH \cite{R5703}, have yielded
$\vel V_{tb} V_{ts}^\ast \ver \sim 0.035$, which is in confirmation with the
CKM estimates. 

Rare semileptonic decays $b\rar s(d) \ell^+ \ell^-$ can provide alternative
sources for searching new physics beyond the SM, and these decays are
relatively clean compared to pure hadronic decays. The matrix elements of the
$b \rar s \ell^+ \ell^-$ transition contain terms describing the virtual
effects induced by the $t\bar{t}$, $c\bar{c}$ and $u\bar{u}$ loops, which
are proportional to $\vel V_{tb} V_{ts}^\ast \ver$, $\vel V_{cb} V_{cs}^\ast
\ver$ and $\vel V_{ub} V_{us}^\ast \ver$, respectively. 

Using the unitarity condition of the CKM matrix and neglecting $\vel V_{ub}
V_{us}^\ast \ver$ in comparison to $\vel V_{tb} V_{ts}^\ast \ver$ and $\vel
V_{cb} V_{cs}^\ast \ver$, it is obvious that, the matrix element for the 
$b\rar s \ell^+ \ell^-$ decay involves only one independent CKM matrix
element, namely, $\vel V_{tb} V_{ts}^\ast \ver$, so that CP--violation in
this channel is strongly suppressed in the SM.

As has already been noted, $b \rar q \ell^+ \ell^-$ decay is a promising
candidate for establishing new physics beyond the SM. New physics effects
manifest themselves in rare B decays in two different ways; either through
new contributions to the Wilson coefficients existing in the SM, or
through the new structures in the effective Hamiltonian, which are absent in
the SM. Note that, the semileptonic $b \rar q \ell^+ \ell^-$ decay has
extensively been studied in numerous works \cite{R5704}--\cite{R5719},
in the framework of the SM and its various extensions. Recently, the first
measurement of the $b \rar s \ell^+ \ell^-$ decay has been reported by BELLE
\cite{R5720} and is in agreement with the SM expectation. Therefore, this
result puts further constraint on any extension of the SM.

In the present work we examine CP violating effects for the case when one of
the leptons is polarized, in model independent framework, by taking into
account a more general form of the effective Hamiltonian. It should be noted
here that similar calculation has been carried out in the SM in \cite{R5721}. 

One efficient way in establishing new physics beyond the SM is the
measurement of the lepton polarization \cite{R5722}--\cite{R5733}. This
issue has been studied for the $b \rar s \tau^+ \tau^-$ transition and 
$B \rar K^\ast \ell^+ \ell^-$, $B \rar K \ell^+ \ell^-$ decays in a model
independent way in \cite{R5727} and \cite{R5732,R5733}, respectively.        

The paper is organized as follows. In section 2, using the most general 
form of the four--Fermi interaction, we derive
model independent expressions for the CP violating asymmetry, for polarized
and unpolarized leptons. In section 3 we present our numerical analysis.

\section{The formalism}

In this section we present the necessary expressions for CP violating
asymmetry when lepton is polarized and unpolarized, using the most general
form of four--Fermi interactions. Following \cite{R5725,R5727}, we write the
matrix element of the $b \rar d \ell^+ \ell^-$ transition in terms of the
twelve model independent four--Fermi interactions

\bea
\label{e5701}
{\cal M} \es \frac{G \alpha}{\sqrt{2} \pi} V_{tb}V_{td}^\ast \Bigg\{
C_{SL} \bar d i \sigma_{\mu\nu} \frac{q^\nu}{q^2} L b \bar \ell \gamma^\mu
\ell + C_{BR} \bar d i \sigma_{\mu\nu} \frac{q^\nu}{q^2} R b \bar \ell
\gamma^\mu \ell + C_{LL}^{tot} \bar d_L \gamma_\mu b_L \bar \ell_L
\gamma^\mu \ell_L \nnb \\
\ar C_{LR}^{tot} \bar d_L \gamma_\mu b_L \bar \ell_R
\gamma^\mu \ell_R + C_{RL} \bar d_R \gamma_\mu b_R \bar \ell_L
\gamma^\mu \ell_L + C_{RR} \bar d_R \gamma_\mu b_R \bar \ell_R
\gamma^\mu \ell_R \nnb \\
\ar C_{LRLR} \bar d_L b_R \bar \ell_L \ell_R +
C_{RLLR} \bar d_R b_L \bar \ell_L \ell_R +
C_{LRRL} \bar d_L b_R \bar \ell_R \ell_L +
C_{RLRL} \bar d_R b_L \bar \ell_R \ell_L \nnb \\
\ar C_T \bar d \sigma_{\mu\nu} b \bar \ell \sigma^{\mu\nu} \ell +
i C_{TE} \epsilon_{\mu\nu\alpha\beta} \bar d \sigma^{\mu\nu} b
\bar \ell \sigma^{\alpha\beta} \ell \Bigg\}~,
\eea
where $L$ and $R$ stand for the chiral operators $L=(1-\gamma_5)/2$ and 
$R=(1+\gamma_5)/2$, respectively, and $C_X$ are the coefficients of the
four--Fermi interactions. The first two terms, $C_{SL}$ and
$C_{BR}$ are the nonlocal Fermi interactions which correspond to 
$-2 m_s C_7^{eff}$ and $-2 m_b C_7^{eff}$ in the SM, respectively.
The next four terms are the vector type interactions with coefficients
$C_{LL}^{tot},~C_{LR}^{tot},~ C_{RL}$ and $C_{RR}$. Two of these 
vector interactions containing the coefficients $C_{LL}^{tot}$ and 
$C_{LR}^{tot}$ do already exist in the SM in 
the forms $C_9^{eff}-C_{10}$ and $C_9^{eff}+C_{10}$, respectively.
Therefore $C_{LL}^{tot}$ and $C_{LR}^{tot}$ can be represented as
\bea
C_{LL}^{tot} \es C_9^{eff}- C_{10} + C_{LL}~, \nnb \\
C_{LR}^{tot} \es C_9^{eff}+ C_{10} + C_{LR}~,\nnb
\eea
where $C_{LL}$ and $C_{LR}$ describe the contributions of new physics. The
following four terms with
coefficients $C_{LRLR},~C_{RLLR},~C_{LRRL}$ and $C_{RLRL}$ describe the
scalar type interactions, and the last two terms with the
coefficients $C_T$ and $C_{TE}$ are the tensor type interactions. It should
be noted here that, in further analysis we will assume that all new
Wilson coefficients are real, as is the case in the SM, while only
$C_9^{eff}$ contains imaginary part and it is parametrized in the following
form
\bea
\label{e5702}
C_9^{eff} = \xi_1 + \lambda_u \xi_2~,
\eea
where
\bea
\lambda_u = \frac{V_{ub} V_{ud}^\ast}{V_{tb} V_{td}^\ast} \nnb~,
\eea
and
\bea
\label{e5703} 
\xi_1 \es 4.128  + 0.138 \omega(\hat{s}) + g(\hat{m}_c,\hat{s})
C_0(\hat{m}_b) 
- \frac{1}{2} g(\hat{m}_d,\hat{s}) (C_3 + C_4) \nnb \\
\ek \frac{1}{2}
g(\hat{m}_b,\hat{s}) (4 C_3 + 4 C_4 + 3C_5 + C_6)
+ \frac{2}{9} (3 C_3 + C_4 + 3C_5 + C_6)~,\nnb \\
\xi_2 \es [g(\hat{m}_c,\hat{s}) - g(\hat{m}_u,\hat{s})](3 C_1 + C_2)~,
\eea 
where $\hat{m}_q = m_q/m_b$, $\hat{s}=q^2$, $C_0(\mu)=3 C_1 + C_2 + 3 C_3 + 
C_4 + 3 C_5 + C_6$, and
\bea
\label{e5704}
\omega(\hat{s}) \es -\frac{2}{9} \pi^2 -\frac{4}{3} Li_2(\hat{s})-
\frac{2}{3} \ln (\hat{s}) \ln(1-\hat{s}) -
\frac{5+4\hat{s}}{3(1+2\hat{s})} \ln(1-\hat{s}) \nnb \\
\ek \frac{2 \hat{s}(1+\hat{s})(1-2\hat{s})}{3(1-\hat{s})^2(1+2\hat{s})}
\ln (\hat{s}) + \frac{5+9 \hat{s}-6 \hat{s}^2}{3(1-\hat{s})(1+2\hat{s})}~,
\eea
represents the $O(\alpha_s)$ correction coming from one gluon exchange
in the matrix element of the operator ${\cal O}_9$ \cite{R5734}, while the 
function $g(\hat{m}_q,\hat{s})$ represents one--loop corrections to the 
four--quark operators $O_1$--$O_6$ \cite{R5735}, whose form is
\bea
\label{e5705}
\lefteqn{
g(\hat{m}_q,\hat{s}) = -\frac{8}{9} \ln (\hat{m}_q) + \frac{8}{27} + 
\frac{4}{9} y_q - \frac{2}{9} (2+y_q)} \nnb \\
\ek \sqrt{\vel 1-y_q \ver} \Bigg\{ \theta(1-y_q)\Bigg[\ln \Bigg(\frac{1+\sqrt{1-y_q}}
{1-\sqrt{1-y_q}}\Bigg) -i\pi \Bigg] + \theta(y_q-1) 
\arctan \Bigg( \frac{1}{\sqrt{y_q-1}}\Bigg) \Bigg\}~,
\eea
where $y_q=4 \hat{m}_q^2/\hat{s}$.

In addition to the short distance contributions, $B\rar X_d \ell^+ \ell^-$
decay also receives long distance contributions, which have their origin in
the real $\bar{u}u$, $\bar{d}d$ and $\bar{c}c$ intermediate states, i.e.,
$\rho$, $\omega$ and $J/\psi$ family. There are four different approaches in
taking long distance contributions into consideration: a) HQET based
approach \cite{R5736}, b) AMM approach \cite{R5737}, c) LSW
approach \cite{R5738}, and d) KS approach \cite{R5733}. In the present work we
choose the AMM approach, in which these resonance contributions are
parametrized using the Breit--Wigner form for the resonant states. The
effective coefficient $C_9^{eff}$ including the $\rho$, $\omega$ and
$J/\psi$ resonances are defined as
\bea
\label{e5706}
C_9^{eff} = C_9(\mu) + Y_{res}(\hat{s})~,
\eea
where
\bea
\label{e5707}
Y_{res} \es -\frac{3\pi}{\alpha^2} \Bigg\{ \ga C^{(0)} (\mu) + \lambda_u
\left[3 C_1(\mu)+ C_2(\mu) \right] \dr 
\sum_{V_i=\psi} K_i \frac{\Gamma(V_i \rar \ell^+\ell^-) M_{V_i}}
{M_{V_i}^2-q^2-iM_{V_i}\Gamma_{V_i}} \nnb \\
\ek \lambda_u g(\hat{m}_u,\hat{s}) \left[3 C_1(\mu)+ C_2(\mu) \right] 
\sum_{V_i=\rho,\omega} \frac{\Gamma(V_i \rar \ell^+\ell^-) M_{V_i}} 
{M_{V_i}^2-q^2-iM_{V_i}\Gamma_{V_i}} \Bigg\}~.
\eea
The phenomenological factor $K_i$ has the universal value for the inclusive 
$B \rar X_s \ell^+ \ell^-$ decay $K_i\simeq 2.3$ \cite{R5739}, which we use 
in our calculations.

As we have already noted, CP asymmetry can appear both for cases when lepton
is polarized or unpolarized, and hence, along this line, we will present the
expressions for the differential decay rate for both cases when the lepton
is polarized and unpolarized. Starting with Eq. (\ref{e5701}), after lengthy
calculations we get the following expression for the unpolarized decay
width:
\bea
\label{e5708}
\frac{d\Gamma}{d\hat{s}} = \frac{G^2 \alpha^2 m_b^5}{32768\pi^5}
\vel V_{tb}V_{td}^\ast \ver^2 \lambda^{1/2}(1,\hat{s},0) v
\Delta(\hat{s})~,
\eea
where $\hat{s}=q^2/m_b^2$, $v=\sqrt{1-4\hat{m}_\ell^2/\hat{s}}$ is the
velocity of the final lepton, $\hat{m}_\ell = m_\ell/m_b$, and
$\lambda(a,b,c)=a^2+b^2+c^2-2ab-2ac-2bc$ is the usual triangle function. The
explicit expression of the function $\Delta(\hat{s})$ can be written as 

\bea
\label{e5709}
\Delta(\hat{s}) \es 16 (1-\hat{s})\, {\rm Re} \Bigg\{
 \frac{8}{3 \hat{s}^2} (2 \hat{m}_\ell^2 +\hat{s} ) (2 + \hat{s})
\Big[ \ve C_{BR} \ve^2 + 8 \hat{s} \ga \ve C_{T}\ve^2
              + 4 \ve C_{TE}\ve^2 \dr \Big]\nnb \\
\ek \frac{8}{\hat{s}} ( 2 \hat{m}_\ell^2 +\hat{s} ) \ga C_{LL}^{tot} + 
C_{LR}^{tot} \dr C_{BR}^\ast
- \frac{32}{\hat{s}} \hat{m}_\ell (2 + \hat{s}) \ga C_{T} - 
2 C_{TE} \dr  C_{BR}^\ast \nnb \\
\ek \frac{4}{3 \hat{s}} [2 \hat{s} (\hat{m}_\ell^2-\hat{s}) - (2 \hat{m}_\ell^2 +\hat{s} )] 
\ga \ve C_{LL}^{tot}\ve^2 + \ve C_{LR}^{tot}\ve^2
+ \ve C_{RL}\ve^2 + \ve C_{RR}\ve^2 \dr \nnb \\
\ek 2 ( 2 \hat{m}_\ell^2-\hat{s}) \ga \ve C_{LRLR}\ve^2
            +  \ve C_{RLLR}\ve^2
            +  \ve C_{LRRL}\ve^2
            +  \ve C_{RLRL}\ve^2 \dr \nnb \\
\ar 8 \hat{m}_\ell^2 \Big[ 2 \ga C_{LL}^{tot} (C_{LR}^{tot})^\ast
                 + C_{RL} C_{RR}^\ast \dr 
- \ga C_{LRLR} C_{LRRL}^\ast
+ C_{RLLR} C_{RLRL}^\ast \dr \Big] \nnb \\
\ek 4 \hat{m}_\ell \Big[ \ga C_{LL}^{tot} - C_{LR}^{tot}\dr \ga
C_{LRLR}^\ast - C_{LRRL}^\ast \dr  + \ga C_{RL} - C_{RR} \dr \ga
C_{RLLR}^\ast - C_{RLRL}^\ast \dr \nnb \\
\ek 12 \ga C_{LL}^{tot} + C_{LR}^{tot} \dr 
\ga C_{T}^\ast - 2 C_{TE}^\ast \dr -
12 \ga C_{RL} + C_{RR} \dr 
\ga C_{T}^\ast + 2  C_{TE}^\ast \dr \Big]
                                     \Bigg\}~,
\eea

Our result agrees with the one given in \cite{R5725}, except
the term multiplying the coefficient $N_9(s)$ in \cite{R5725}. The differential 
decay width for the CP conjugated process can be obtained from Eq.
(\ref{e5707}) 
by making the replacement  
\bea
\Delta \rar \bar{\Delta},~\mbox{\rm i.e.,}~C_9^{eff} = \xi_1+\lambda_u \xi_2 \rar 
\bar{C}_9^{eff}=\xi_1 + \lambda_u^\ast \xi_2~.\nnb
\eea

The lepton polarization has been firstly analyzed in the SM in \cite{R5733}
and \cite{R5740}, where it has been shown that additional information can be
obtained about the quadratic functions of the Wilson coefficients
$C_7^{eff}$, $C_9^{eff}$ and $C_{10}$. In order to calculate the final lepton
polarization, we define the orthogonal unit vectors $\vec{e}_L$,
$\vec{e}_T$ and $\vec{e}_N$ in such a way that, in the rest frame of leptons
they can be expressed as
\bea
s^{-\mu}_L \es \ga 0,\vec{e}_L^{\,-}\dr = 
\ga 0,\frac{\vec{p}_-}{\vel\vec{p}_- \ver}\dr~, \nnb \\    
s^{-\mu}_N \es \ga 0,\vec{e}_N^{\,-}\dr = \ga 0,\frac{\vec{p}_s\times 
\vec{p}_-}{\vel \vec{p}_s\times \vec{p}_- \ver}\dr~, \nnb \\
s^{-\mu}_T \es \ga 0,\vec{e}_T^{\,-}\dr = \ga 0,\vec{e}_N^{\,-} 
\times \vec{e}_L^{\,-} \dr~, \nnb \\
s^{+\mu}_L \es \ga 0,\vec{e}_L^{\,+}\dr = 
\ga 0,\frac{\vec{p}_+}{\vel\vec{p}_+ \ver}\dr~, \nnb \\
s^{+\mu}_N \es \ga 0,\vec{e}_N^{\,+}\dr = \ga 0,\frac{\vec{p}_s\times 
\vec{p}_+}{\vel \vec{p}_s\times \vec{p}_+ \ver}\dr~, \nnb \\
s^{+\mu}_T \es \ga 0,\vec{e}_T^{\,+}\dr = \ga 0,\vec{e}_N^{\,+} 
\times \vec{e}_L^{\,+}\dr~, \nnb
\eea 
where $\vec{p}_-$, $\vec{p}_+$ and $\vec{p}_s$ are the three--momenta of the
leptons $\ell^-$, $\ell^+$, and the strange quark in the
center of mass frame (CM) of $\ell^- \,\ell^+$, respectively, and the subscripts
$L$, $N$ and $T$ stand for the longitudinal, normal and transversal
polarization of the lepton. Boosting the unit vectors $s^-_L$ and $s^+_L$ 
corresponding to longitudinal polarization by Lorentz transformation, from
the rest frame of the corresponding leptons, to the $\ell^- \,\ell^+$ CM 
frame, we get
\bea
\label{e5710}
\ga s^{-\mu}_L \dr_{CM} \es \ga \frac{\vel\vec{p}_- \ver}{m_\ell}~,
\frac{E \vec{p}_-}{m_\ell \vel\vec{p}_- \ver}\dr~,\nnb \\
\ga s^{+\mu}_L \dr_{CM} \es \ga \frac{\vel\vec{p}_- \ver}{m_\ell}~,   
-\frac{E \vec{p}_-}{m_\ell \vel\vec{p}_- \ver}\dr~,  
\eea
while $s^{\mp\mu}_N$ and $s^{\mp\mu}_T$ are not changed by the boost.

The differential decay rate of the $b \rar d \ell^+ \ell^-$ decay, for
any spin direction $\vec{n}^{\mp}$ of $\ell^\mp$, where $\vec{n}^\mp$ is the
unit vector in the rest frame of $\ell^\mp$, can be written in the following
form
\bea
\label{e5711}
\frac{d\Gamma(s,\vec{n}^\mp)}{d\hat{s}} =
\frac{1}{2} \frac{d\Gamma}{d\hat{s}}_0 \left[1+
\ga P_L^\mp \vec{e}_L^{\,\mp} + P_N^\mp \vec{e}_N^{\,\mp}+
P_T^\mp \vec{e}_T^{\,\mp}\dr \cdot \vec{n}^{\mp} \right]~,
\eea
where $(d\Gamma/d\hat{s})_0$ corresponds to the unpolarized
differential decay width (see Eq. (\ref{e5708})) for the $b \rar d \ell^+
\ell^-$ decay. The differential decay width for the $\bar{b} \rar \bar{d} \ell^+
\ell^-$ decay, can simply be obtained by making the replacement
\bea
\frac{d\Gamma(s,\vec{n}^\mp)}{d\hat{s}} \rar
\frac{d\bar{\Gamma}(s,\vec{n}^\mp)}{d\hat{s}}~, \nnb
\eea
where $d\bar{\Gamma}/d\hat{s}$ is obtained from $d\Gamma/d\hat{s}$ by
replacing $C_9^{eff}=\xi_1+\lambda_u \xi_2$ to
$\bar{C}_9^{eff}=\xi_1+\lambda_u^\ast \xi_2$.
The polarizations $P_L$, $P_N$ and $P_T$ are defined as
\bea
\label{e5712}
P_i^\mp(\hat{s}) = \frac{\ds{\frac{d\Gamma}{d\hat{s}}(\vec{n}^\mp=\vec{e}_i^{\,\mp})}-
\ds{\frac{d\Gamma}{d\hat{s}}(\vec{n}^\mp=-\vec{e}_i^{\,\mp})}}
{\ds{\frac{d\Gamma}{d\hat{s}}(\vec{n}^\mp=\vec{e}_i^{\,\mp})}+ 
\ds{\frac{d\Gamma}{d\hat{s}}(\vec{n}^\mp=-\vec{e}_i^{\,\mp})}} =
\frac{\Delta_i^\mp}{\Delta}~,
\eea
with $i=L,~N,~T$.

The explicit expressions for the longitudinal polarization asymmetries 
$P_L^-$ and $P_L^+$ are

\bea
P_L^- \es  \frac{32 (1-\hat{s}) v}{\Delta}\, {\rm Re} \Bigg\{
4 \ga C_{LL}^{tot}- C_{LR}^{tot} \dr C_{BR}^\ast \nnb \\
\ek \frac{2}{3} (1 + 2 \hat{s}) \ga \ve C_{LL}^{tot}\ve^2 - 
\ve C_{LR}^{tot}\ve ^2 + 
\ve C_{RL}\ve ^2 - \ve C_{RR}\ve ^2 - 128 C_{T} C_{TE}^\ast \dr \nnb \\
\ar \frac{16}{3 \hat{s}} \hat{m}_\ell (2 + \hat{s}) 
\ga  C_{T} - 2 C_{TE} \dr C_{BR}^\ast
+ 2 \hat{m}_\ell \Big[\ga C_{LL}^{tot} - C_{LR}^{tot}\dr 
\ga  C_{LRLR}^* + C_{LRRL}^* \dr \nnb \\
\ar \ga C_{RL} - C_{RR}\dr \ga  C_{RLLR}^* + C_{RLRL}^* \dr 
- 4  \ga 3 C_{LL}^{tot} -C_{LR}^{tot}\dr \ga C_T^*  - 2 C_{TE}^* \dr \nnb \\
\ek 4\ga  C_{RL} - 3 C_{RR} \dr \ga C_T^*  + 2 C_{TE}^* \dr \Big]
- \hat{s} \Big( \ve C_{LRLR}\ve ^2 - \ve C_{LRRL}\ve ^2 \nnb \\
\ar \ve C_{RLLR}\ve ^2 - \ve C_{RLRL}\ve ^2 + 128 C_T^\ast C_{TE} \Big)
\Bigg\}, \nnb
\eea

and

\bea
P_L^+ \es \frac{32 (1-\hat{s}) v}{\Delta}\, {\rm Re}  \Bigg\{
- 4 \ga \bar{C}_{LL}^{tot}- \bar{C}_{LR}^{tot} \dr 
C_{BR}^\ast  \nnb \\
\ar\frac{2}{3}(1 + 2 \hat{s}) \ga  \ve \bar{C}_{LL}^{tot}\ve ^2 - 
\ve \bar{C}_{LR}^{tot}\ve ^2 + \ve C_{RL}\ve ^2 - \ve C_{RR}\ve ^2
+ 128 C_{T} C_{TE}^\ast \dr   \nnb \\
\ar\frac{16}{3 \hat{s}} \hat{m}_\ell (2 + \hat{s}) \ga C_{T} - 
2 C_{TE} \dr  C_{BR}^\ast
+ 2 \hat{m}_\ell \Big[ \ga \bar{C}_{LL}^{tot} - \bar{C}_{LR}^{tot}\dr 
\ga C_{LRLR}^* + C_{LRRL}^* \dr \nnb \\
\ar \ga C_{RL} - C_{RR}\dr \ga C_{RLLR}^* + C_{RLRL}^* \dr
+ 4  \ga  \bar{C}_{LL}^{tot} - 3 \bar{C}_{LR}^{tot}\dr 
\ga C_T^*  - 2 C_{TE}^* \dr \nnb \\
\ar 4 \ga 3 C_{RL} - C_{RR} \dr \ga C_T^*  + 2 C_{TE}^* \dr \Big]
- \hat{s} \Big( \ve C_{LRLR}\ve ^2 - \ve C_{LRRL}\ve ^2 \nnb \\
\ar \ve C_{RLLR}\ve ^2 - \ve C_{RLRL}\ve ^2 + 128 C_T^\ast C_{TE} \Big)
\Bigg\}. \nnb
\eea

The normal asymmetries, $P_N^-$ and $P_N^+$, are;

\bea
P_N^- \es - \frac{4 \pi (1-\hat{s}) v \sqrt{\hat{s}}}{\Delta} \, {\rm Im} \Bigg\{
- \frac{8}{\hat{s}} \hat{m}_\ell \ga C_{LL}^{tot} - 
C_{LR}^{tot} \dr C_{BR}^\ast \nnb \\
\ar 8 \hat{m}_\ell \Big[ C_{LL}^{tot} (C_{LR}^{tot})^\ast - C_{RL} C_{RR}^\ast
- 2 \ga C_{LRLR} + C_{LRRL}\dr \ga C_{T}^\ast - 2 C_{TE}^\ast \dr\nnb \\
\ek 2 \ga C_{RLLR} + C_{RLRL} \dr \ga C_{T}^\ast + 2 C_{TE}^\ast \dr\Big]
+ 4 \Big[ \ga C_{LRLR} + C_{LRRL} \dr C_{BR}^\ast \nnb \\
\ar  C_{LL}^{tot} C_{LRRL}^\ast +  C_{LR}^{tot} C_{LRLR}^\ast
+  C_{RL} C_{RLRL}^\ast +  C_{RR} C_{RLLR}^\ast
- 4  C_{RL} \ga  C_{T}^\ast + 2 C_{TE}^\ast \dr \nnb \\
\ek 4 C_{LR}^{tot} \ga  C_{T}^\ast - 2 C_{TE}^\ast \dr \Big]
- \frac{16}{\hat{s}} \ga  C_{T} -2 C_{TE} \dr C_{BR}^\ast 
            \Bigg\}~, \nnb 
\eea

\bea
P_N^+ \es - \frac{4 \pi (1-\hat{s}) v \sqrt{\hat{s}}}{\Delta} \, {\rm Im} \Bigg\{
\frac{8}{\hat{s}} \hat{m}_\ell  \ga \bar{C}_{LL}^{tot} - \bar{C}_{LR}^{tot}
\dr C_{BR}^\ast \nnb \\
\ar 8 \hat{m}_\ell \Big[ - \bar{C}_{LL}^{tot} (\bar{C}_{LR}^{tot})^\ast + C_{RL} C_{RR}^\ast
+ 2 \ga C_{LRLR} + C_{LRRL}\dr \ga C_{T}^\ast - 2 C_{TE}^\ast \dr \nnb \\
\ar 2 \ga C_{RLLR} + C_{RLRL} \dr \ga C_{T}^\ast + 2 C_{TE}^\ast \dr \Big]
- 4 \Big[ \ga C_{LRLR} + C_{LRRL} \dr C_{BR}^\ast \nnb \\
\ar  \bar{C}_{LL}^{tot} C_{LRLR}^\ast
+  \bar{C}_{LR}^{tot} C_{LRRL}^\ast
+  C_{RL} C_{RLLR}^\ast 
+  C_{RR} C_{RLRL}^\ast
+ 4  \bar{C}_{LL}^{tot} \ga  C_{T}^\ast - 2 C_{TE}^\ast \dr \nnb \\
\ar 4  C_{RR} \ga  C_{T}^\ast + 2 C_{TE}^\ast \dr \Big]
+ \frac{16}{\hat{s}} C_{BR} \ga  C_{T}^\ast -2 C_{TE}^\ast \dr 
\Bigg\}~.\nnb
\eea

The transverse asymmetries, $P_T^-$ and $P_T^+$, are;

\bea
P_T^- \es - \frac{8 \pi (1-\hat{s})}{\Delta \sqrt{\hat{s}}}
   \, {\rm Re} \Bigg\{
- \frac{8}{\hat{s}} \hat{m}_\ell \ve C_{BR}\ve ^2
+ 4 \hat{m}_\ell \ga 3 C_{LL}^{tot}+  C_{LR}^{tot}\dr C_{BR}^\ast  \nnb \\
\ek 2 ( 1 +\hat{s}) \hat{m}_\ell \ga \ve C_{LL}^{tot}\ve ^2 - \ve C_{RR}\ve ^2\dr 
+ 4 \hat{m}_\ell \hat{s} \Big[- C_{LL}^{tot} (C_{LR}^{tot})^\ast + C_{RL} C_{RR}^\ast \nnb \\
\ar 2 \ga  C_{LRLR} - C_{LRRL} \dr \ga C_{T}^\ast - 2 C_{TE}^\ast \dr 
+ 2 \ga  C_{RLLR} - C_{RLRL} \dr \ga  C_{T}^\ast + 2 C_{TE}^\ast \dr \Big] \nnb \\
\ar 2 (1 - \hat{s}) \hat{m}_\ell \ga \ve C_{LR}^{tot}\ve ^2 - \ve C_{RL}\ve ^2 \dr
- 2 \hat{s}  \ga  C_{LRLR} - C_{LRRL} \dr C_{BR}^\ast \nnb \\
\ar \frac{8}{\hat{s}} ( 4 \hat{m}_\ell^2 + \hat{s} ) C_{BR} \ga C_{T}^\ast - 2 C_{TE}^\ast\dr 
+ 4 \hat{m}_\ell^2 \Big[ C_{LL}^{tot} C_{LRLR}^\ast - 
C_{LR}^{tot} C_{LRRL}^\ast \nnb \\
\ar C_{RL} C_{RLLR}^\ast - C_{RR} C_{RLRL}^\ast 
- 12 C_{LL}^{tot} \ga C_T^\ast - 2 C_{TE}^\ast \dr + 12
C_{RR} \ga C_T^\ast + 2 C_{TE}^\ast \dr \Big] \nnb \\
\ar 2 ( 2 \hat{m}_\ell^2 - \hat{s} ) \Big[ C_{LL}^{tot} C_{LRRL}^\ast - 
C_{LR}^{tot} C_{LRLR}^\ast 
+ C_{RL} C_{RLRL}^\ast - C_{RR} C_{RLLR}^\ast  \nnb \\
\ar 4 C_{LR}^{tot} \ga C_T^\ast - 2 C_{TE}^\ast \dr -
4 C_{RL} \ga C_T^\ast + 2 C_{TE}^\ast \dr \Big]
+ 256 \hat{m}_\ell C_{T} C_{TE}^\ast \Bigg\}~,\nnb
\eea

and

\bea
P_T^+ \es - \frac{8 \pi (1-\hat{s})}{\Delta \sqrt{\hat{s}}}
   \, {\rm Re} \Bigg\{
- \frac{8}{\hat{s}} \hat{m}_\ell \ve C_{BR}\ve ^2
+ 4 \hat{m}_\ell \ga \bar{C}_{LL}^{tot}+ 3 \bar{C}_{LR}^{tot}\dr C_{BR}^\ast  \nnb \\
\ar 2 (1 - \hat{s}) \hat{m}_\ell \ga \ve \bar{C}_{LL}^{tot}\ve ^2 - \ve C_{RR}\ve ^2\dr 
- 4 \hat{m}_\ell \hat{s} \Big[\bar{C}_{LL}^{tot} (\bar{C}_{LR}^{tot})^\ast - C_{RL} C_{RR}^\ast \nnb \\
\ar 2 \ga  C_{LRLR} - C_{LRRL}\dr \ga C_{T}^\ast - 2C_{TE}^\ast \dr 
+ 2 \ga  C_{RLLR} - C_{RLRL} \dr \ga  C_{T}^\ast + 2 C_{TE}^\ast \dr \Big] \nnb \\
\ek 2 (1 +\hat{s}) \hat{m}_\ell \ga \ve \bar{C}_{LR}^{tot}\ve ^2 - \ve C_{RL}\ve ^2 \dr
+ 2 \hat{s}  \ga  C_{LRLR} - C_{LRRL} \dr C_{BR}^\ast \nnb \\
\ar \frac{8}{\hat{s}} (4 \hat{m}_\ell^2 +\hat{s}) C_{BR} \ga C_{T}^\ast - 2 C_{TE}^\ast\dr 
+ 4 \hat{m}_\ell^2 \Big[ \bar{C}_{LL}^{tot} C_{LRRL}^\ast - 
\bar{C}_{LR}^{tot} C_{LRLR}^\ast \nnb \\
\ar C_{RL} C_{RLRL}^\ast - C_{RR} C_{RLLR}^\ast 
-12 \bar{C}_{LR}^{tot} \ga C_{T}^\ast - 2 C_{TE}^\ast \dr
+12 C_{RL} \ga C_{T}^\ast + 2 C_{TE}^\ast \dr \Big] \nnb \\
\ar 2 ( 2 \hat{m}_\ell^2 - \hat{s} ) \Big[ \bar{C}_{LL}^{tot} C_{LRLR}^\ast - 
\bar{C}_{LR}^{tot} C_{LRRL}^\ast 
+ C_{RL} C_{RLLR}^\ast - C_{RR} C_{RLRL}^\ast \nnb \\
\ar 4 \bar{C}_{LL}^{tot} \ga C_{T}^\ast - 2 C_{TE}^\ast \dr
- 4 C_{RR} \ga C_{T}^\ast + 2 C_{TE}^\ast \dr \Big]
+ 256 \hat{m}_\ell C_{T} C_{TE}^\ast \Bigg\}~.\nnb
\eea

It should be noted here that, these polarizations were calculated in
\cite{R5725}, using the most general form of the effective Hamiltonian.
Our results for $P_L$ and $P_N$ agree with the ones given in \cite{R5725},
while the transversal polarizations $P_T^-$ and $P_T^+$ both differ from the
ones given in the same work. In the SM case, our results for $P_L$, $P_N$
and $P_T$ coincide with the results of \cite{R5721}.
It is quite obvious from the expressions of $P_i$ that, they involve
various quadratic combinations of the Wilson coefficients and hence they are
quite sensitive to the new physics. The polarizations $P_N$ and $P_T$ are
proportional to $m_\ell$ and therefore can be significant for $\tau$ lepton
only. 

Having obtained all necessary expressions, we can proceed now to study 
the CP violating asymmetries. In the unpolarized lepton case, the CP
violating differential decay width asymmetry is defined as
\bea
\label{e5713}
A_{CP}(\hat{s}) = \frac{\ds{\ga \frac{d\Gamma}{d\hat{s}}\dr_0}-
\ds{\ga \frac{d\bar{\Gamma}}{d\hat{s}}\dr_0} }
{\ds{\ga \frac{d\Gamma}{d\hat{s}}\dr_0}+
\ds{\ga \frac{d\bar{\Gamma}}{d\hat{s}}\dr_0} } =
\frac{\Delta -\bar{\Delta}}{\Delta +\bar{\Delta}}~,
\eea
where
\bea
\frac{d\Gamma}{d\hat{s}} = \frac{d\Gamma(b\rar
d\ell^+\ell^-)}{d\hat{s}},~\mbox{\rm and},~
\frac{d\bar{\Gamma}}{d\hat{s}} = \frac{d\bar{\Gamma}(b\rar
d\ell^+\ell^-)}{d\hat{s}}~,\nnb
\eea
and $(d\bar{\Gamma}/d\hat{s})_0$ can be obtained from
$(d\Gamma/d\hat{s})_0$ by making the replacement
\bea
\label{e5714}
C_9^{eff} = \xi_1+\lambda_u\xi_2 \rar \bar{C}_9^{eff}=
\xi_1+\lambda_u^\ast\xi_2~.
\eea

Using Eqs. (\ref{e5711}) and (\ref{e5713}), we get for the CP violating asymmetry
\bea
\label{e5715}
A_{CP}(\hat{s}) \es - 4 \mbox{\rm Im}[\lambda_u]
\frac{\Sigma(\hat{s})}{\Delta(\hat{s})+\bar{\Delta}(\hat{s})}~,\nnb\\
\ap - 2 \mbox{\rm Im}[\lambda_u] \frac{\Sigma(\hat{s})}{\Delta(\hat{s})}~,
\eea
and $\Sigma(\hat{s})$, whose explicit form we do not present, can easily be
obtained using Eqs. (\ref{e5709}) and (\ref{e5713}).

In the presence of the lepton polarization CP asymmetry is modified and
the source of this modification can be attributed to the presence of new
interference terms which contain $C_9^{eff}$ (in our case $C_{LL}^{tot}$ and
$C_{LR}^{tot}$). We now proceed to calculate this new contribution. 

In the polarized lepton case, CP asymmetry can be defined as
\bea
\label{e5716}
A_{CP}(\vec{s}) =
\frac{\ds{\frac{d\Gamma}{d\hat{s}}(\hat{s},\vec{n}^-)}-
\ds{\frac{d\bar{\Gamma}}{d\hat{s}}(\hat{s},\vec{n}^+)}}
{\ds{\ga \frac{d\Gamma}{d\hat{s}}\dr_0}+
\ds{\ga \frac{d\bar{\Gamma}}{d\hat{s}}\dr_0} }~,
\eea
where
\bea
\frac{d\Gamma}{d\hat{s}} = \frac{d\Gamma(b\rar
d\ell^+\ell^-(\vec{n}^-))}{d\hat{s}},~\mbox{\rm and},~
\frac{d\bar{\Gamma}}{d\hat{s}} = \frac{d\Gamma(b\rar
d\ell^+(\vec{n}^+)\ell^-)}{d\hat{s}}~.\nnb  
\eea

The differential decay width with lepton polarization for the 
$b \rar d \ell^+ \ell^-$ channel is given by Eq. (\ref{e5711}). Analogously,
for the corresponding CP conjugated process we have the expression
\bea
\label{e5717}
\frac{d\bar{\Gamma}}{d\hat{s}}(\vec{n}^\mp) = \frac{1}{2} \ga
\frac{d\bar{\Gamma}}{d\hat{s}}\dr_0 \left[ 1 + P_i^+ (\vec{e}_i^{\,\mp}
\cdot \vec{n}^\mp) \right]~.
\eea               
With the choice $\vec{n}^-=\vec{n}^+$, $P_i^+$ can be constructed from the
differential decay width analogous to Eq. (\ref{e5712}).   
At this stage we have all necessary ingredients for calculation of the CP violating
asymmetry for the lepton $\ell^-$ with polarization $\vec{n}=\vec{e}_i$.
Inserting Eqs. (\ref{e5711}) and (\ref{e5717}) into Eq. (\ref{e5716}), and setting
$\vec{n}^-=\vec{n}^+$, the CP violating asymmetry when lepton is polarized
is defined as
\bea
A_{CP} \es
\frac{\ds{\frac{1}{2}\ga \frac{d\Gamma}{d\hat{s}}\dr_0 
\left[ 1+P_i^- ( \vec{e}_i^{\, -}\cdot \vec{n} ) \right]} -
\ds{\frac{1}{2}\ga \frac{d\bar{\Gamma}}{d\hat{s}}\dr_0 
\left[ 1+P_i^+ ( \vec{e}_i^{\, +}\cdot \vec{n} ) \right]}}
{\ds{\ga \frac{d\Gamma}{d\hat{s}}\dr_0} +                                            
 \ds{\ga \frac{d\bar{\Gamma}}{d\hat{s}}\dr_0}}~. \nnb
\eea
Taking into account the fact that $\vec{e}_{L,N}^{\, +}= -\vec{e}_{L,N}^{\, -}$, and
$\vec{e}_{T}^{\, +}= \vec{e}_{T}^{\, -}$, we obtain
\bea
\label{e5718}
A_{CP} \es \frac{1}{2} \left\{
\frac{\ds{\ga \frac{d\Gamma}{d\hat{s}}\dr_0} -
 \ds{\ga \frac{d\bar{\Gamma}}{d\hat{s}}\dr_0}}
{\ds{\ga \frac{d\Gamma}{d\hat{s}}\dr_0} +
 \ds{\ga \frac{d\bar{\Gamma}}{d\hat{s}}\dr_0}} 
\pm
\frac{\ds{\ga \frac{d\Gamma}{d\hat{s}}\dr_0
P_i^-} \mp
\ds{\ga \frac{d\bar{\Gamma}}{d\hat{s}}\dr_0
P_i^+}}
{\ds{\ga \frac{d\Gamma}{d\hat{s}}\dr_0} +
 \ds{\ga \frac{d\bar{\Gamma}}{d\hat{s}}\dr_0}}
\right\}~.
\eea
Using Eq. (\ref{e5708}), we get from Eq. (\ref{e5718}),
\bea
\label{e5719}
A_{CP} (\vec{n} = \mp \vec{e}_i^{\, -})\es
\frac{1}{2} \left\{ \frac{\Delta-\bar{\Delta}}
{\Delta+\bar{\Delta}} \pm 
\frac{\Delta_i\mp\bar{\Delta}_i}{\Delta+\bar{\Delta}}\right\}~, \nnb \\
\es \frac{1}{2} \left\{A_{CP} (\hat{s}) \pm \delta A_{CP}^i (\hat{s})
\right\}~,
\eea
where the upper sign in the definition of $\delta A_{CP}$ corresponds to $L$
and $N$ polarizations, while the lower sign corresponds to $T$ polarization.
  
The $\delta A_{CP}^i (\hat{s})$ terms in Eq. (\ref{e5719}) describe the
modification to the unpolarized decay width, which can be written as
\bea
\label{e5720}
\delta A_{CP}^i (\hat{s}) \es \frac {-4 \mbox{\rm Im} \lambda_u \delta
\Sigma^i(\hat{s})}{\Delta(\hat{s})+\bar{\Delta}(\hat{s})}~, \nnb \\
\ap -2 \mbox{\rm Im} \lambda_u \frac{\delta
\Sigma^i(\hat{s})}{\Delta(\hat{s})}~.
\eea
We do not present the explicit forms of the expressions for 
$\delta\Sigma^i(\hat{s}),~(i=L,N,T)$, since their calculations are 
straightforward.

\section{Numerical analysis}

In this section we will study the dependence of the CP asymmetries 
$A_{CP}(\hat{s})$ and $\delta A_{CP}^i (\hat{s})$ on $\hat{s}$ at fixed
values of the new Wilson coefficients. Once again, we remind the reader that
in the present work all new Wilson coefficients are taken to be real. The
experimental result on $b \rar s \gamma$ decay put strong restriction on
$C_{BR}$, i.e., practically it has the same value as it has in the SM.
Therefore, in further numerical analysis we will set $C_{BR}=-2 C_7^{eff}$.
Throughout numerical analysis, we will vary all new Wilson coefficients in
the range $-4 \le C_X \le 4$. The experimental bounds on the branching
ratios of the $B \rar K(K^\ast) \mu^+ \mu^-$ \cite{R5741,R5742} and 
$B \rar \mu^+ \mu^-$ \cite{R5743} suggest that this is the right order of 
magnitude for vector and scalar type interactions. Recently, BaBar and BELLE
collaborations \cite{R5740,R5741} have presented new results on the
branching ratios of $B \rar K \ell^+ \ell^-$ and $B \rar K^\ast \ell^+ \ell^-$
decays. When these results are used, stronger restrictions are imposed on
some of the new Wilson coefficients. For example, $-2 \le C_{LL} \le 0$,
$0 \le C_{RL} \le 2.3$, $-1.5 \le C_{T} \le 1.5$ and $-3.3 \le C_{TE} \le
2.6$, and rest of the coefficients vary in the region $-4 \le C_X \le
4$. However, since the results of BaBar and BELLE are preliminary we will
not take into account these results in the present analysis and vary all of
the new Wilson coefficients in the region $\-4 \le C_X \le 4$. Furthermore,
in our analysis we will use the Wolfenstein parametrization \cite{R5744} for
the CKM matrix. The currently allowed range for the Wolfenstein parameters are:
$0.19 \le \rho \le 0.268$ and $0.19 \le \eta \le 0.268$ \cite{R5745}, where
in the present analysis they are set to $\rho=0.25$ and $\eta=0.34$.   

We start our numerical analysis by first discussing the dependence of
$A_{CP}$ on $\hat{s}$ at fixed values of $C_i$, i.e., $C_i = -4;0;4$ which
can be summarized as follows

\begin{itemize}
\item 
For the $b \rar d \mu_+ \mu^-$ case, far from resonance
regions, $A_{CP}$ depends strongly on $C_{LL}$. We observe
that, when $C_{LL}$ is positive (negative), the value of $A_{CP}$
decreases (increases), while the situation for the $C_{LR}$ case is the
opposite way around (see Figs. (1) and (2)). If the tensor interaction is
taken into account, $A_{CP}$ practically seems to be zero for all values of
$C_T$ and $C_{TE}$. Furthermore, $A_{CP}$ shows quite a weak dependence on
all remaining Wilson coefficients and the departure from the SM result is
very small.

\item 
For the $b \rar d \tau_+ \tau^-$ case, in the region between
second and third $\psi$ resonances, $A_{CP}$ is sensitive to
$C_{LR}$, $C_{LRLR}$, $C_{LRRL}$, $C_{T}$, and $C_{TE}$, as can be seen from
the Figs (3), (4), (5), (6) and (7), respectively. When $C_{LR}$ and $C_{LRLR}$
are positive (negative), they contribute destructively (constructively) to
the SM result. The situation is contrary to this behavior for the $C_{LRRL}$ 
scalar coupling. In the tensor interaction case, in the second and third
resonance region, the magnitude of $A_{CP}$ is smaller compared to that of
the SM result. But, it is quite important to observe that $A_{CP}$ asymmetry 
changes its sign, compared to its behavior in the SM, when $C_T$ $(C_{TE})$ 
is negative (positive). Therefore, determination of the sign and magnitude of
$A_{CP}$ can give promising information about new physics.
\end{itemize}

The results concerning $\delta A_{CP}$ for the $b \rar d \mu^+ \mu^-$ decay
can be summarized as follows:

\begin{itemize}
\item 
In the region $1~GeV^2/m_b^2 \le \hat{s} \le 8~GeV^2/m_b^2$, which is free of
resonance contribution, CP asymmetry due to the longitudinal polarization of
$\mu$ lepton is dependent strongly on $C_{LL}$, and is practically independent 
of all remaining vector interaction coefficients. When $C_{LL}$ is negative
(positive), $\delta A_{CP}$ is larger (smaller) compared to that of the SM
result (see Fig. (8)). 

\item 
$\delta A_{CP}^L$ depends strongly on all scalar type interactions. As an
example we present the dependence of $\delta A_{CP}$ on $C_{LRRL}$ in Fig. (9). 
The terms proportional to tensor interaction terms contribute destructively to 
the SM result.
\end{itemize}

For the $b \rar d \tau^+ \tau^-$ case, we obtain the following results:
\begin{itemize}
\item
$\delta A_{CP}^L$ depends strongly on the tensor type interactions and when
$C_T$ is negative (positive) it constructive (destructive) contribution to
the SM result. For the other tensor interaction coefficient $C_{TE}$, 
the situation is contrary to this behavior (see Figs. (10) and (11)).

\item
Similar to the $\mu$ lepton case, $\delta A_{CP}^L$ is quite sensitive to the
existence of all scalar type interaction coefficients. When the signs of the
coefficients $C_{LRRL}$ and $C_{LRLR}$ are negative (positive) the sign of
$\delta A_{CP}^L$ is positive (negative), in the region $\hat{s} \ge 0.6$
(see Figs. (12) and (13)).
Note that in the SM case the sign of $\delta A_{CP}^L$ can be positive or
negative. Therefore in the region $\hat{s} \ge 0.6$, determination of
the sign of $\delta A_{CP}^L$ can give unambiguous information about the
existence of new physics beyond the SM.
For the remaining two scalar interaction coefficients $C_{RLRL}$
($C_{RLLR}$), the sign of $\delta A_{CP}^L$ is negative (positive)
(see Figs. (14) and (15)). Again,
as in the previous case, determination of the sign and magnitude of $\delta
A_{CP}^L$ can give quite valuable hints for establishing new physics beyond
the SM.     
\end{itemize}

Since transversal and normal polarizations are proportional to the lepton
mass, for the light lepton case, obviously, departure from the SM results is
not substantial for all Wilson coefficients. On the other hand, for the $b
\rar d \ell^- \ell^+$ transition, $\delta A_{CP}^i~(i=T$ or $N)$ is strongly
dependent on $C_{LR}$ (see Fig. (16)), $C_{RR}$ (see Fig. (17)) and
scalar type interactions. Note that, the dependence of $\delta A_{CP}^T$ on 
$C_T$ and $C_{TE}$ is quite weak.

Finally, we would like to discuss the following question. As has already
been mentioned, $A_{CP}$, as well as $\delta A_{CP}$, are very sensitive to
the existence of new physics beyond the SM. The intriguing question is that,
can we find a region of $C_X$, in which $A_{CP}$ agrees with the SM result
while $\delta A_{CP}$ does not. A possible existence of such a region will
allow us to establish new physics by only measuring $\delta A_{CP}$. In
order to verify whether such a region of $C_X$ does exist or not, we present
the correlations between partially integrated $A_{CP}$ and $\delta A_{CP}$
in Figs. (18)--(20). The integration region for the $b \rar d \mu^+
\mu^-$ transition is chosen to be $1~GeV^2/m_b^2 \le \hat{s} \le 8~GeV^2/m_b^2$, 
and for the $b \rar d \tau^+ \tau^-$ transition it is chosen as
$18~GeV^2/m_b^2 \le \hat{s} \le 1$. These choices of the regions are dictated by
the requirement that $A_{CP}$ and $\delta A_{CP}$ be free of resonance
contributions.      

In Figs. (18)--(19) we present the correlations $\delta A_{CP}^i$ and   
$A_{CP}^i$ asymmetries, when one of the leptons is
longitudinally polarized, for the $\mu$ and $\tau$ lepton cases,
respectively. In Fig. (20) we present the flows in the $(A_{CP}^T$
and $\delta A_{CP}^T)$ plane, when the final lepton is transversally
polarized. From these figures we observe that, indeed, there exists a region
of new Wilson coefficients in which $A_{CP}$ agrees with the SM
prediction, while $\delta A_{CP}$ does not (in Figs (18)--(20),
intersection point of all curves correspond to the SM case).

The numerical values of $\delta A_{CP}^N$ and $A_{CP}^N$ are very small and
for this reason we do not present this correlation. As a final remark we
would like to comment that, similar calculation can be carried out for the
$B \rar \rho \ell^+ \ell^-$ decay in search of new physics, since its
detection in the experiments is easier compared to that of the inclusive $b
\rar d \ell^+ \ell^-$ decay.

In conclusion, we study the CP violating asymmetries, when one the final leptons
is polarized, using the most general form of effective Hamiltonian. It is
seen that, $\delta A_{CP}$ and $A_{CP}$ are very sensitive to various
new Wilson coefficients. Moreover, we discuss the possibility
whether there exist regions of new Wilson coefficients or not, for which
$A_{CP}$ coincides with the SM prediction, while $\delta A_{CP}$ does not.
In other words, if there exists such regions of $C_X$, this means new
physics effects can only be established in $\delta A_{CP}$ measurements. Our
results confirm that, such regions of $C_X$ do indeed exist.

\newpage

\section*{Figure captions}
{\bf Fig. (1)} The dependence of $A_{CP}$ on $\hat{s}$ for the $b \rar d
\mu^+ \mu^-$ transition, at fixed values of 
$C_{LL}$.\\ \\
{\bf Fig. (2)} The same as in Fig. (1), but at fixed values of
$C_{LR}$.\\ \\
{\bf Fig. (3)} The same as in Fig. (1), but for the 
$b \rar d \tau^+ \tau^-$ transition, at fixed values of $C_{LR}$.\\ \\
{\bf Fig. (4)} The same as in Fig. (3), but at fixed values of
$C_{LRRL}$.\\ \\
{\bf Fig. (5)} The same as in Fig. (3), but at fixed values of
$C_{LRLR}$.\\ \\
{\bf Fig. (6)} The same as in Fig. (3), but at fixed values of
$C_{T}$.\\ \\
{\bf Fig. (7)} The same as in Fig. (3), but at fixed values of
$C_{TE}$.\\ \\
{\bf Fig. (8)} The dependence of $\delta A_{CP}^L$ on $\hat{s}$
for the $b \rar d \mu^+ \mu^-$ transition, at fixed values of 
$C_{LL}$, when one of the final leptons is longitudinally 
polarized.\\ \\
{\bf Fig. (9)} The same as in Fig. (8), but at fixed values of
$C_{LRRL}$.\\ \\
{\bf Fig. (10)} The same as in Fig. (8), but for the 
$b \rar d \tau^+ \tau^-$ transition, at fixed values of $C_{T}$.\\ \\
{\bf Fig. (11)} The same as in Fig. (10), but at fixed values of 
$C_{TE}$.\\ \\
{\bf Fig. (12)} The same as in Fig. (10), but at fixed values of 
$C_{LRRL}$.\\ \\
{\bf Fig. (13)} The same as in Fig. (10), but at fixed values of 
$C_{LRLR}$.\\ \\
{\bf Fig. (14)} The same as in Fig. (10), but at fixed values of 
$C_{RLRL}$.\\ \\
{\bf Fig. (15)} The same as in Fig. (10), but at fixed values of 
$C_{RLLR}$.\\ \\
{\bf Fig. (16)} The same as in Fig. (10), but when one of the final leptons
is transversally polarized, at fixed values of $C_{LR}$.\\ \\
{\bf Fig. (17)} The same as in Fig. (16), but at fixed values of 
$C_{RR}$.\\ \\
{\bf Fig. (18)} Parametric plot of the correlation between the partially
integrated $A_{CP}^L$ and $\delta A_{CP}^L$ as a function of the new Wilson
coefficients $C_X$, for the $b \rar d \mu^+ \mu^-$ transition,
when one of the final leptons is longitudinally polarized.\\ \\
{\bf Fig. (19)} The same as in Fig. (18), but for the 
$b \rar d \tau^+ \tau^-$ transition.\\ \\
{\bf Fig. (20)} The same as in Fig. (19), but when one of the final leptons 
is transversally polarized.\\ \\

\newpage

\begin{figure}
\vskip 1.5 cm
    \includegraphics{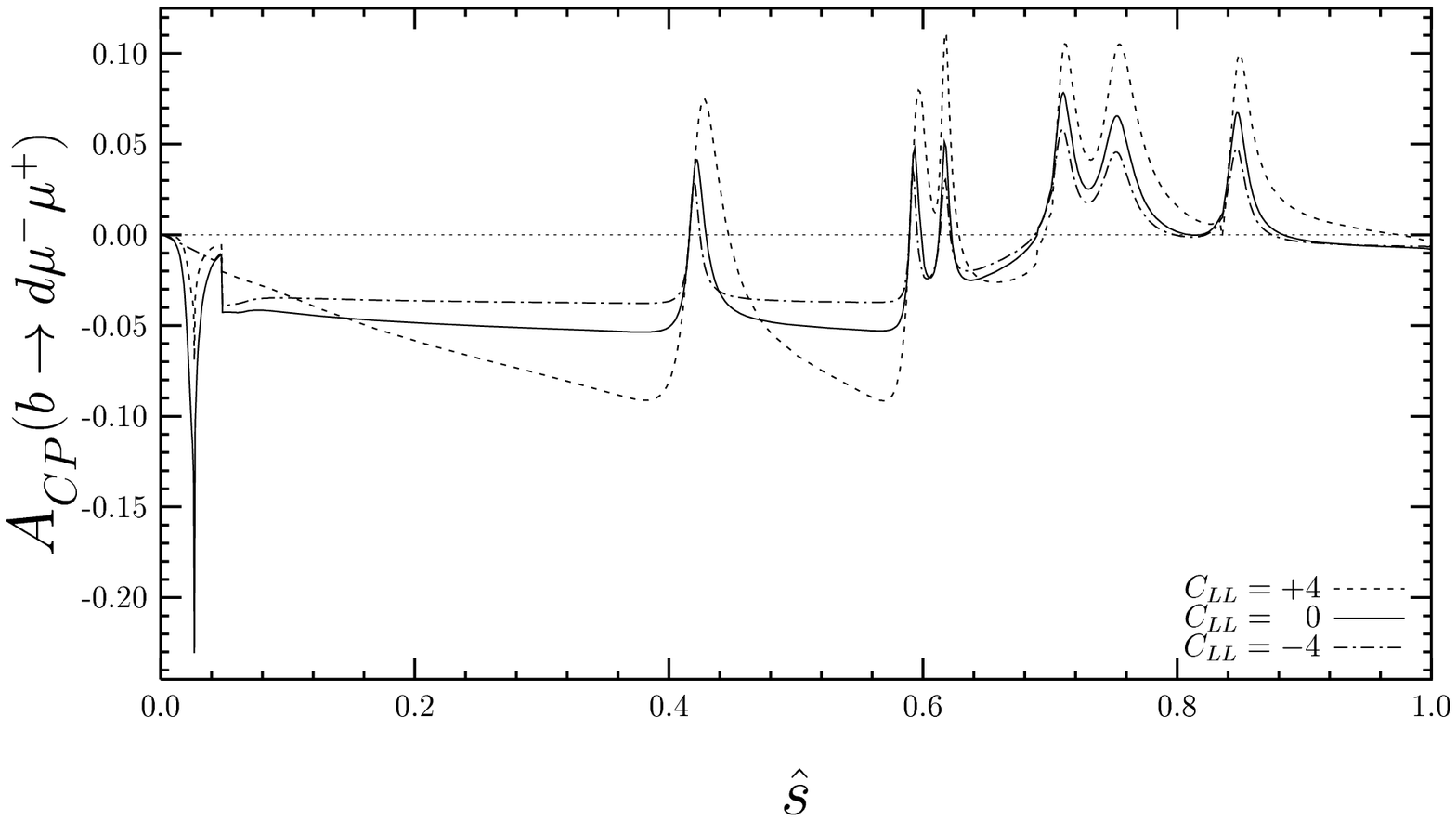}
\vskip 7.8cm
\caption{}
\end{figure}

\begin{figure}
\vskip 2.5 cm
    \includegraphics{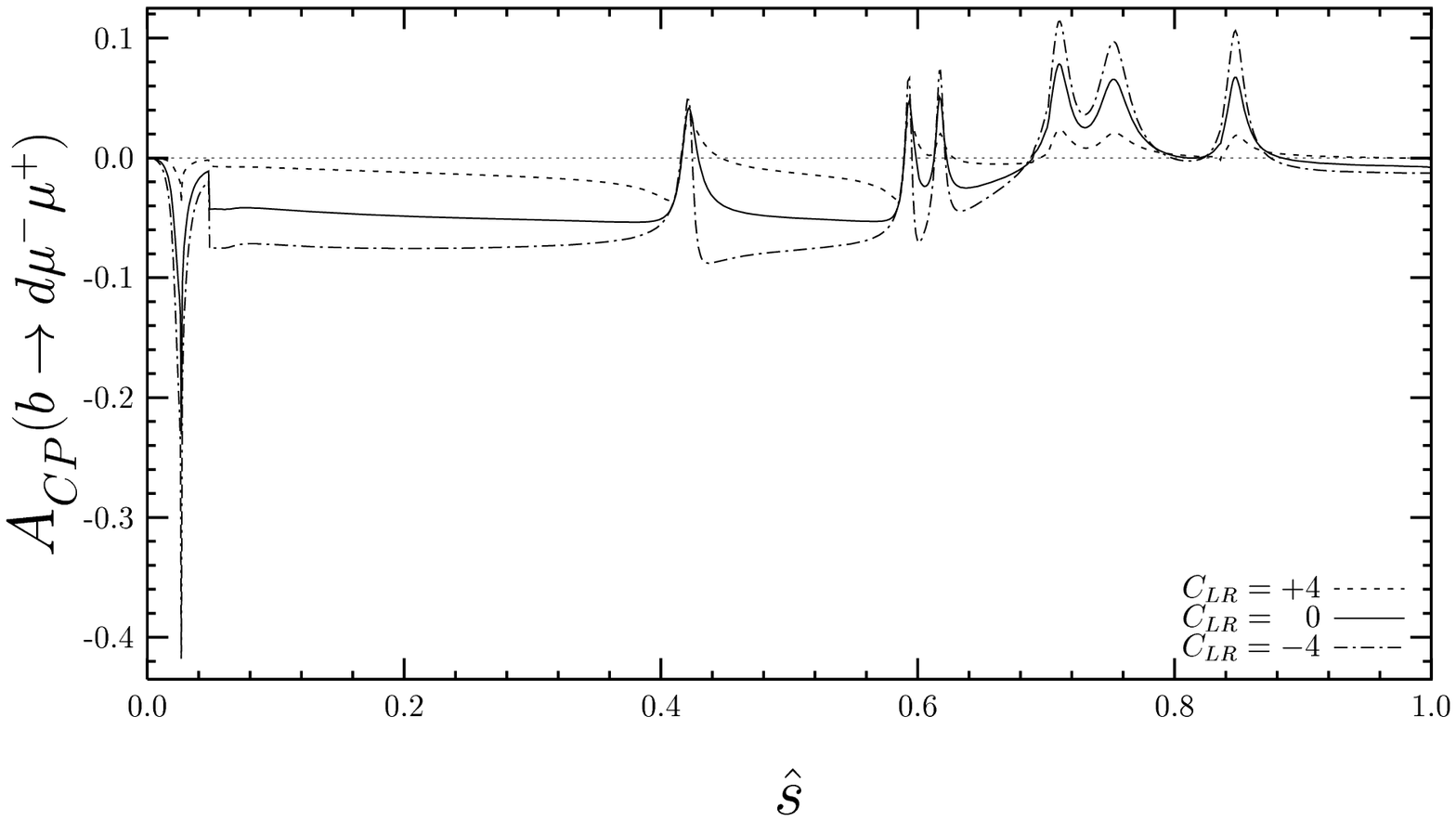}
\vskip 7.8 cm
\caption{}
\end{figure}

\begin{figure}
\vskip 1.5 cm
    \includegraphics{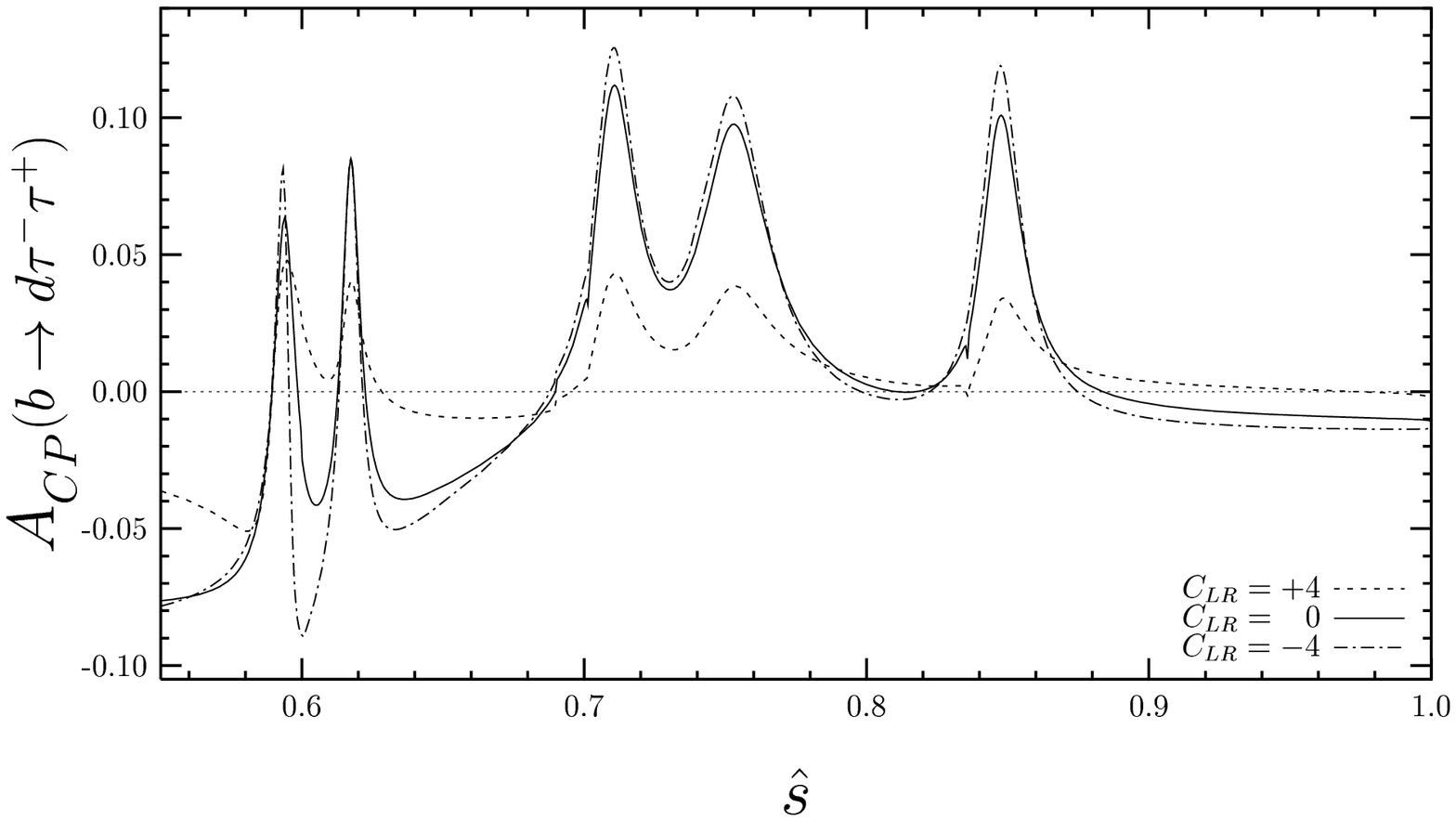}
\vskip 7.8cm
\caption{}
\end{figure}

\begin{figure}
\vskip 2.5 cm
    \includegraphics{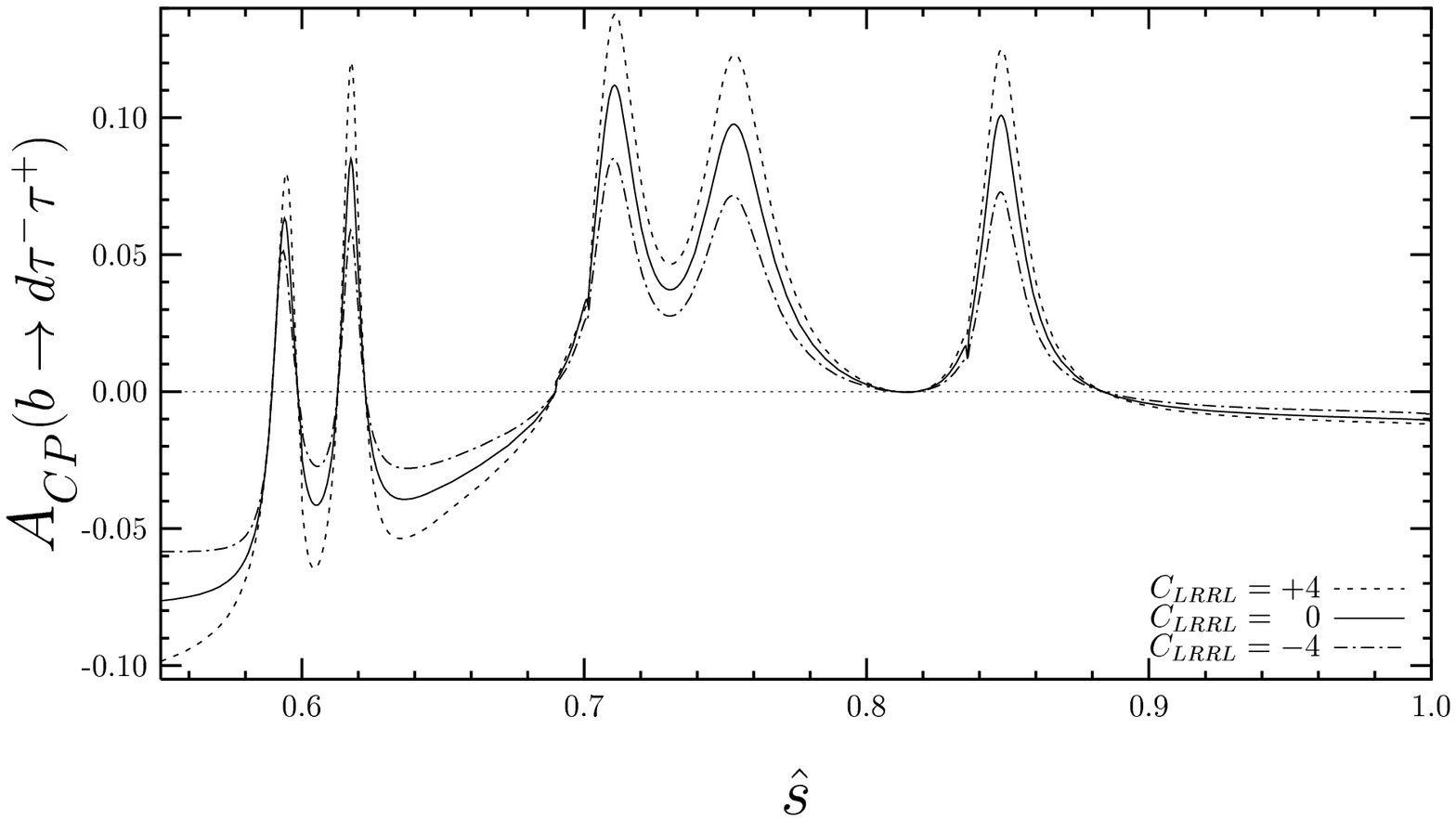}
\vskip 7.8 cm
\caption{}
\end{figure}

\begin{figure}
\vskip 1.5 cm
    \includegraphics{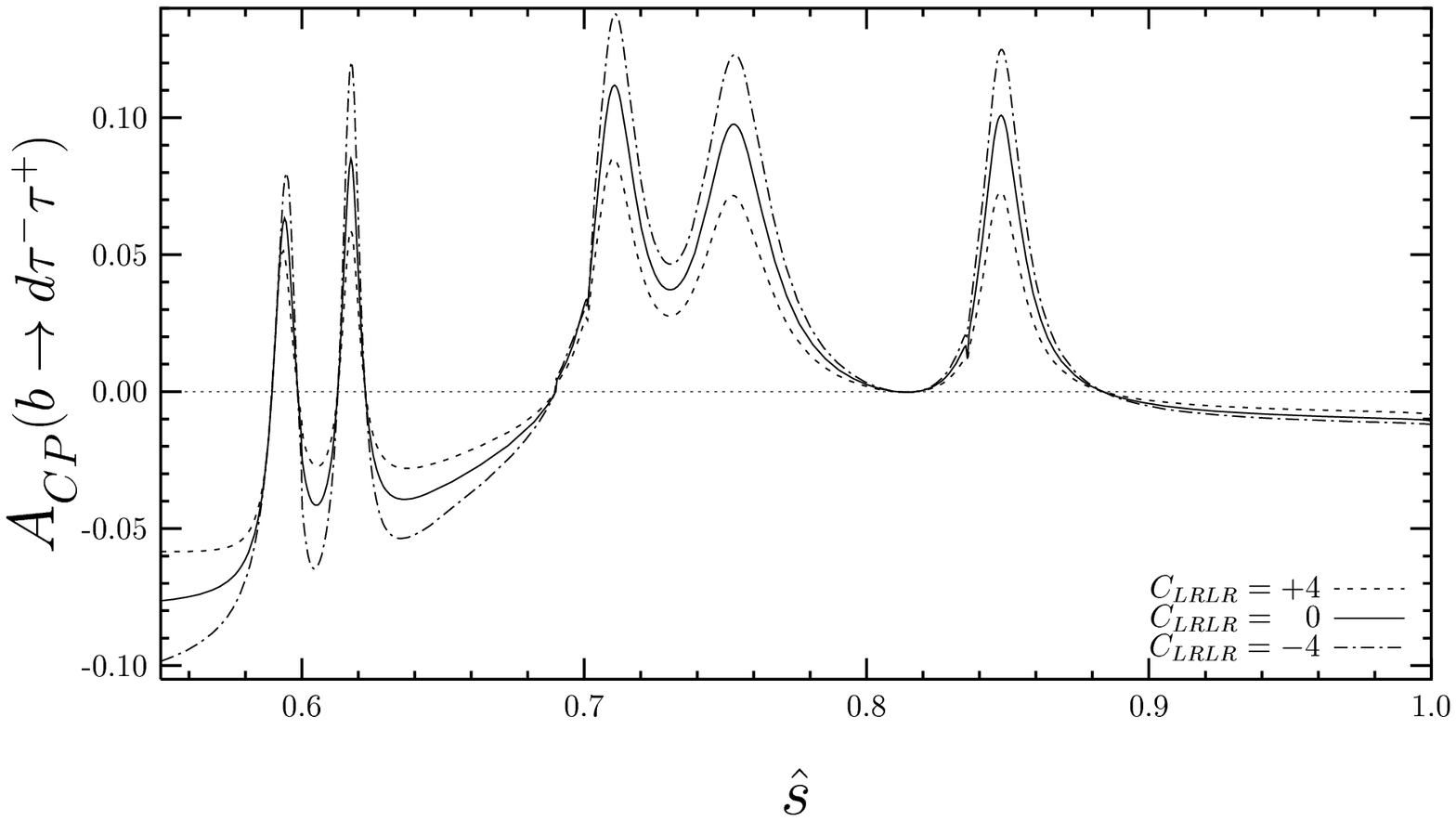}
\vskip 7.8cm
\caption{}
\end{figure}

\begin{figure}
\vskip 2.5 cm
    \includegraphics{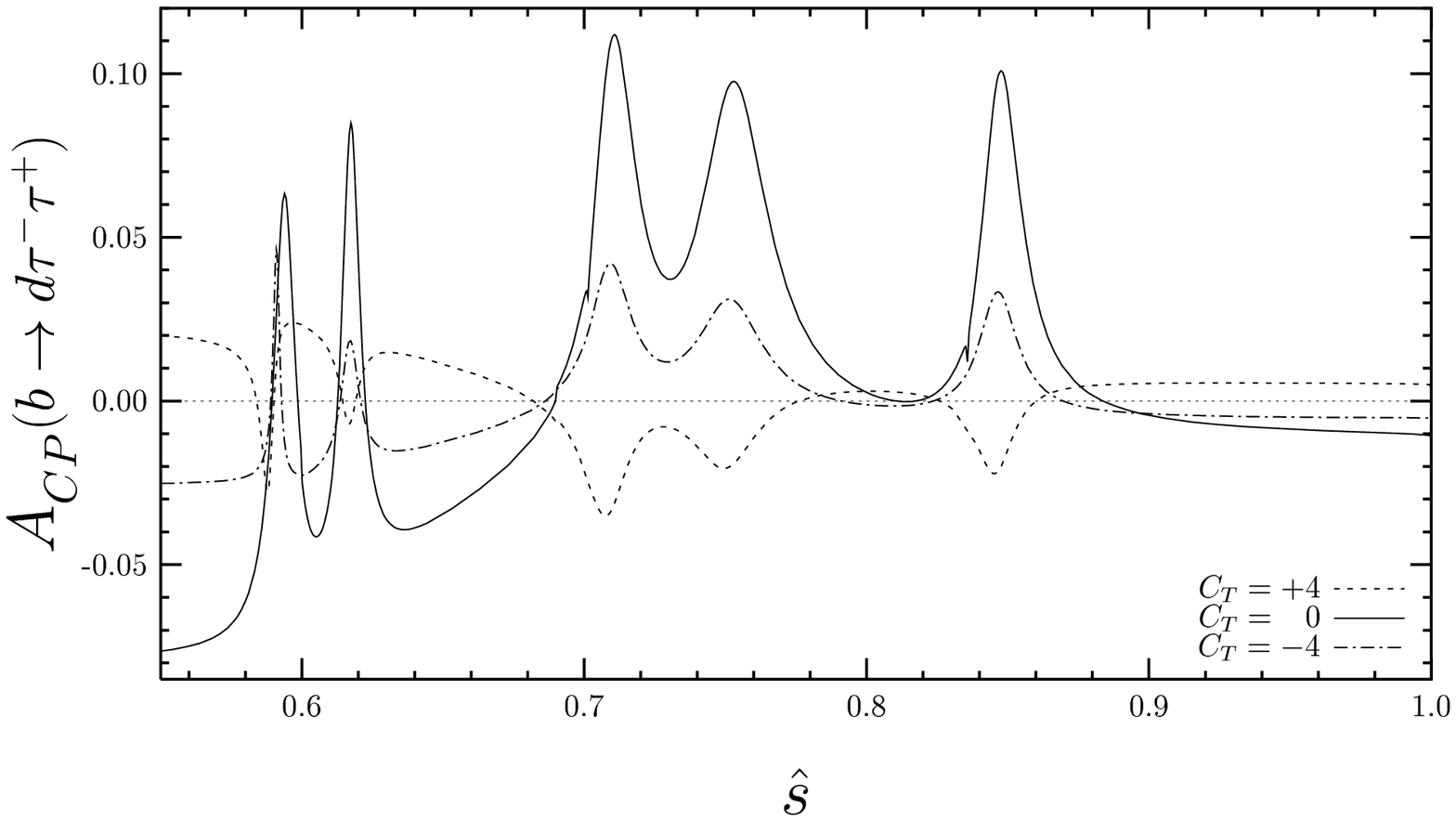}
\vskip 7.8 cm
\caption{}
\end{figure}

\begin{figure}
\vskip 1.5 cm
    \includegraphics{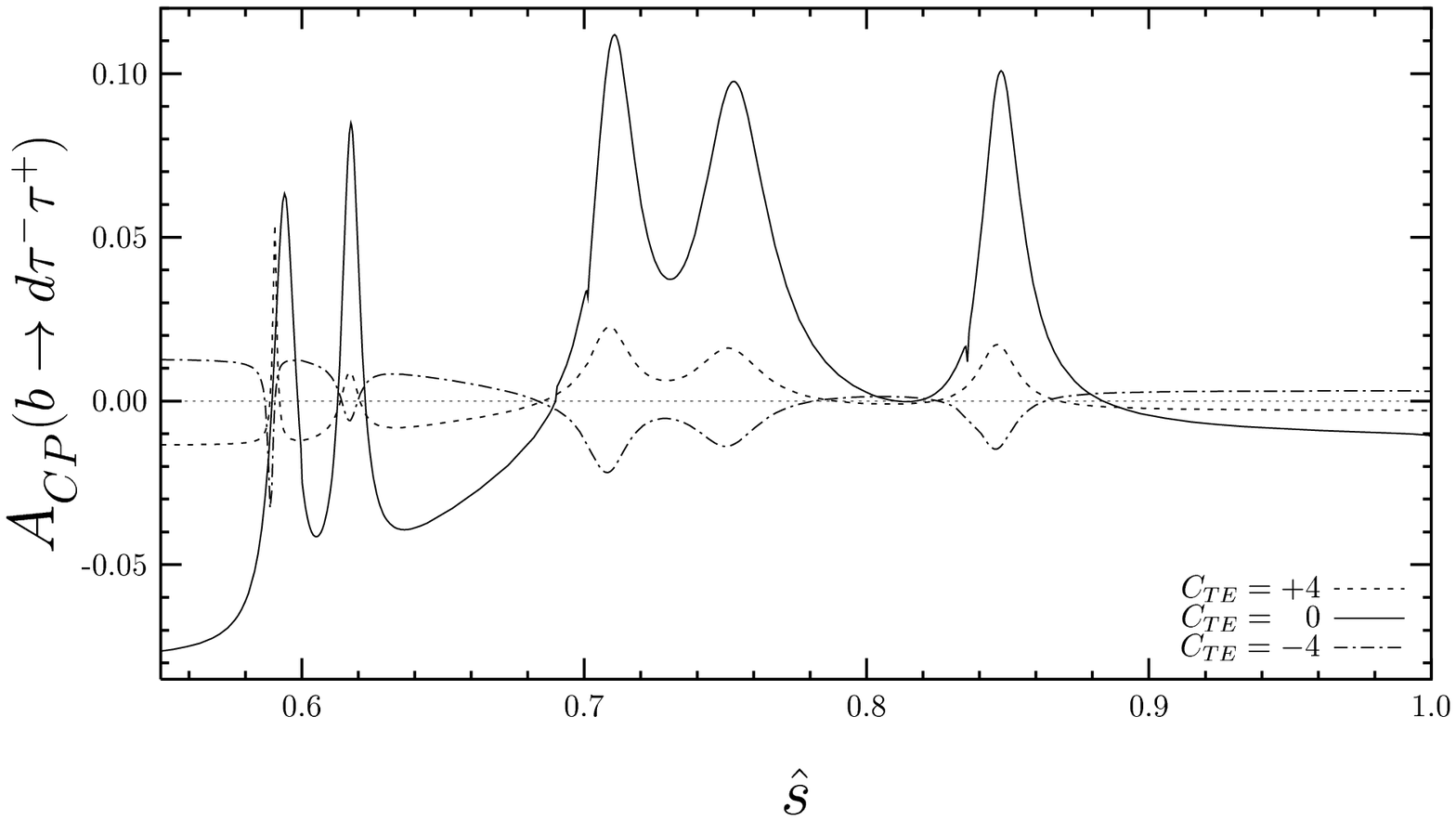}
\vskip 7.8cm
\caption{}
\end{figure}

\begin{figure}
\vskip 2.5 cm
    \includegraphics{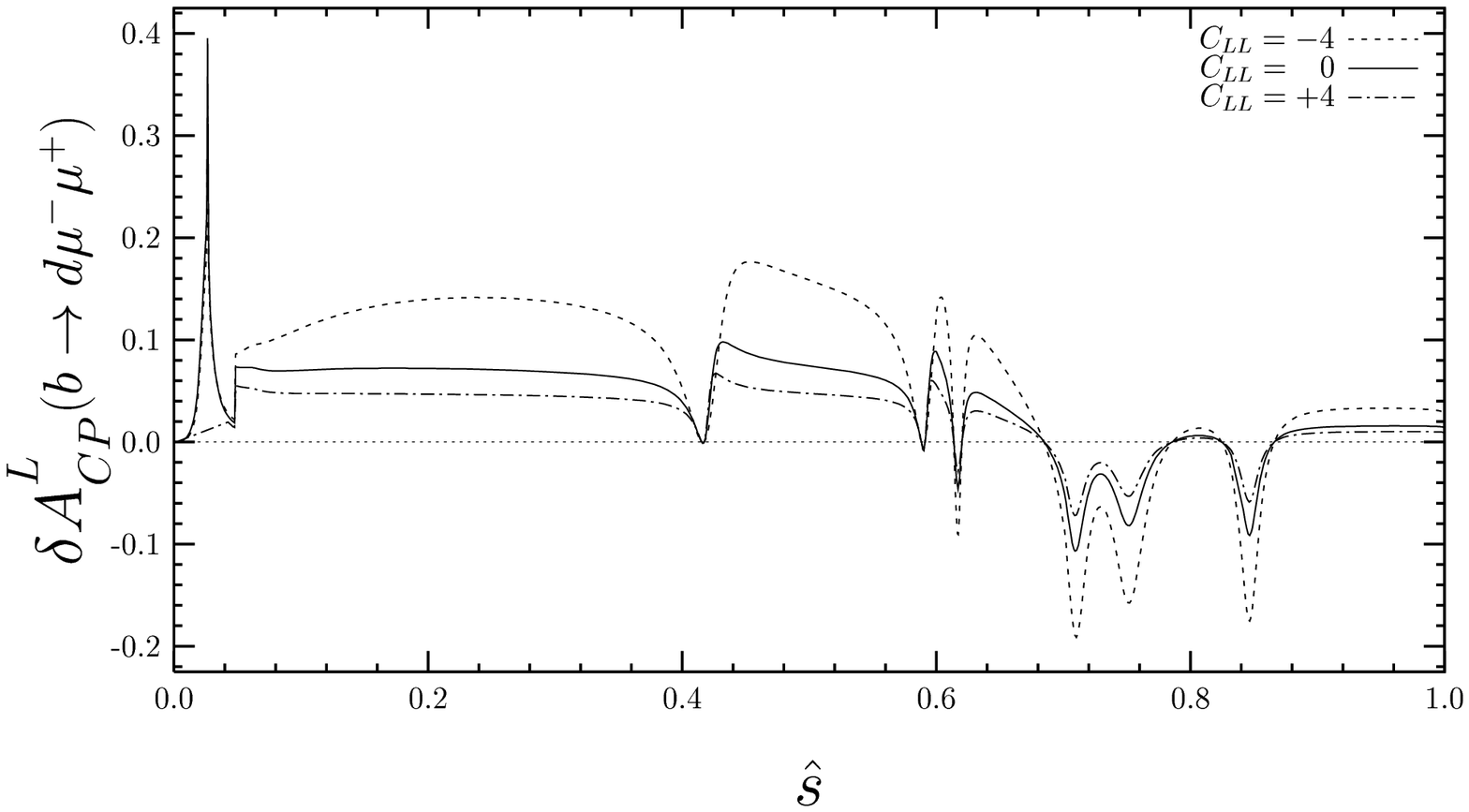}
\vskip 7.8 cm
\caption{}
\end{figure}

\begin{figure}
\vskip 1.5 cm
    \includegraphics{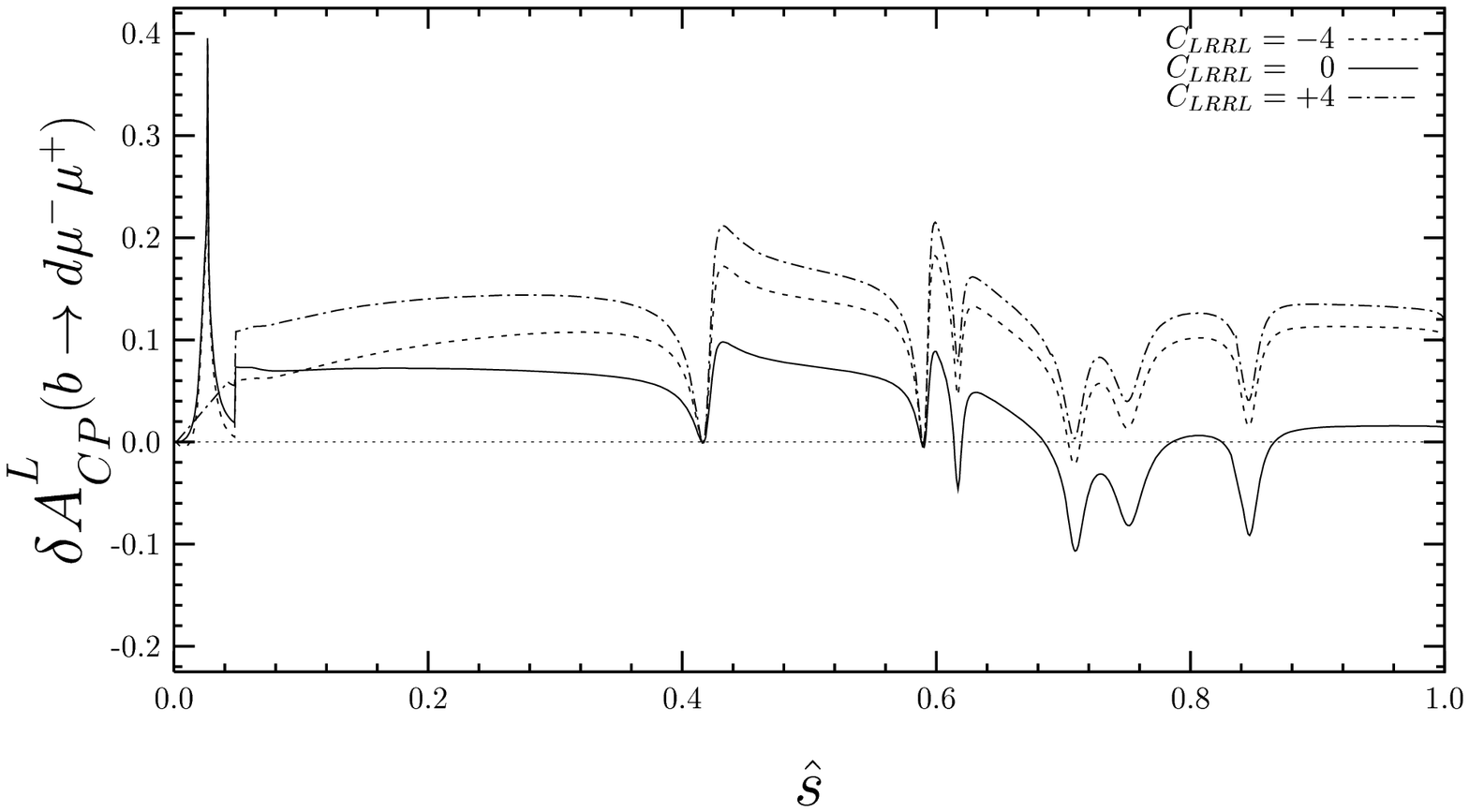}
\vskip 7.8cm
\caption{}
\end{figure}

\begin{figure}
\vskip 2.5 cm
    \includegraphics{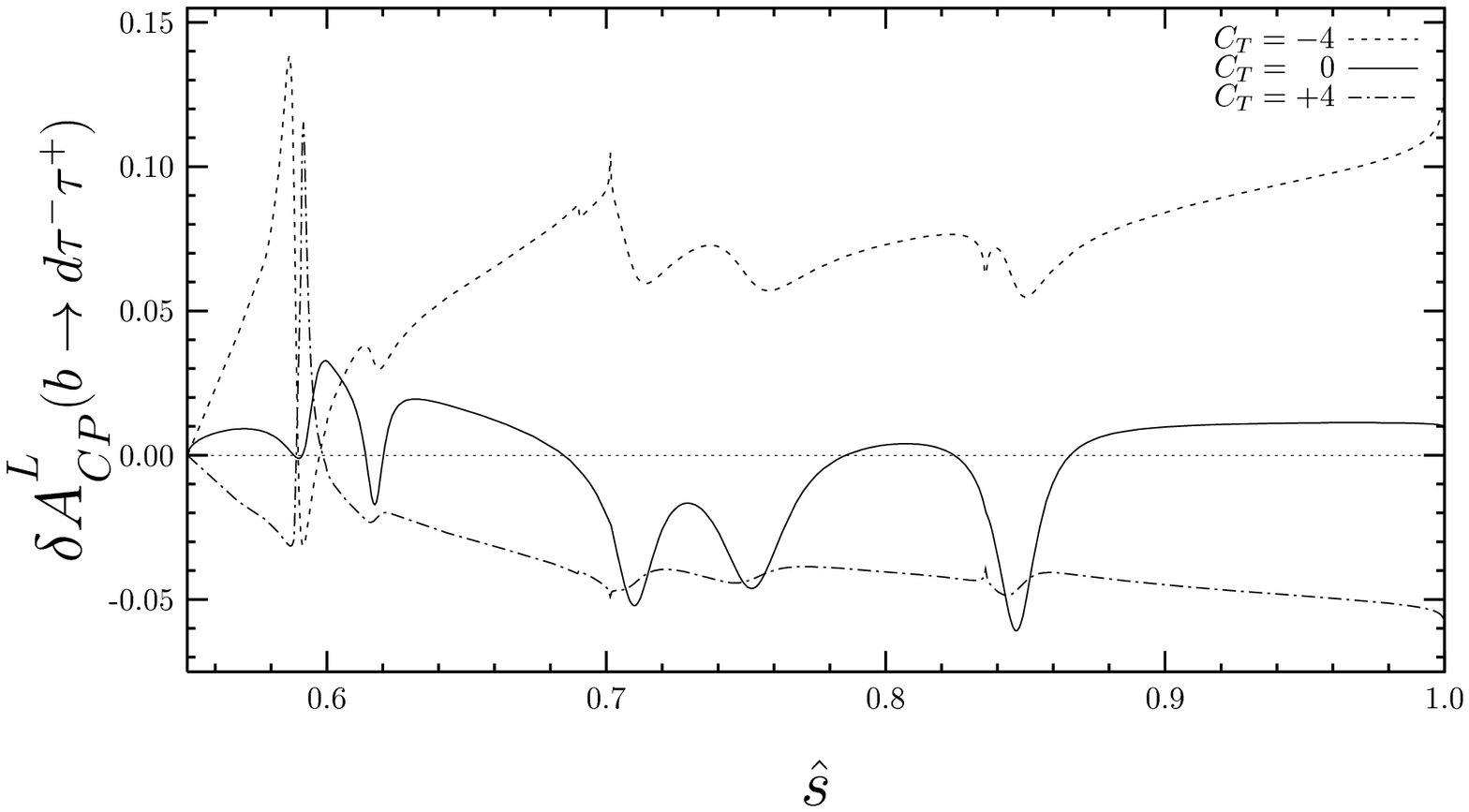}
\vskip 7.8 cm
\caption{}
\end{figure}

\begin{figure}
\vskip 1.5 cm
    \includegraphics{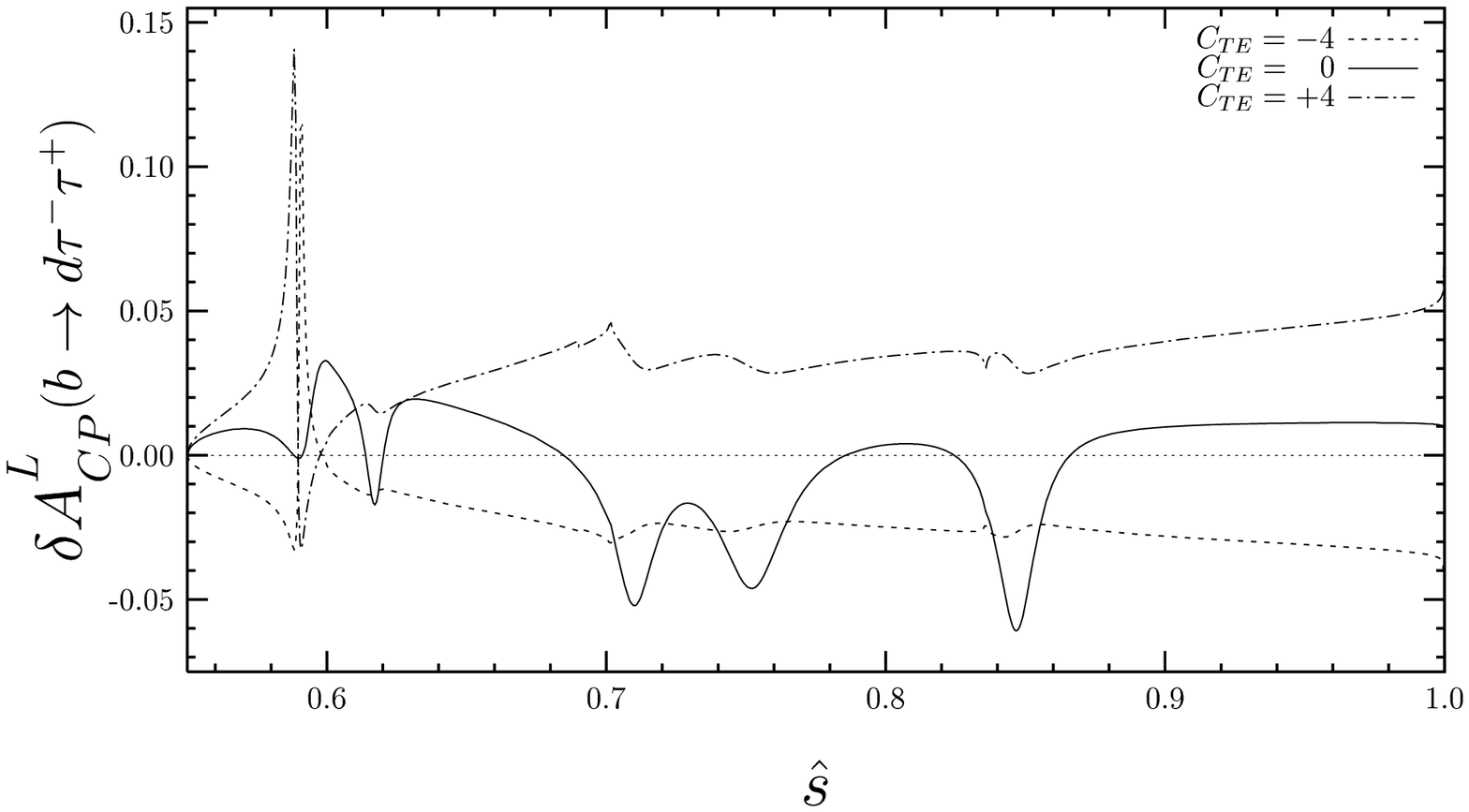}
\vskip 7.8cm
\caption{}
\end{figure}

\begin{figure}
\vskip 2.5 cm
    \includegraphics{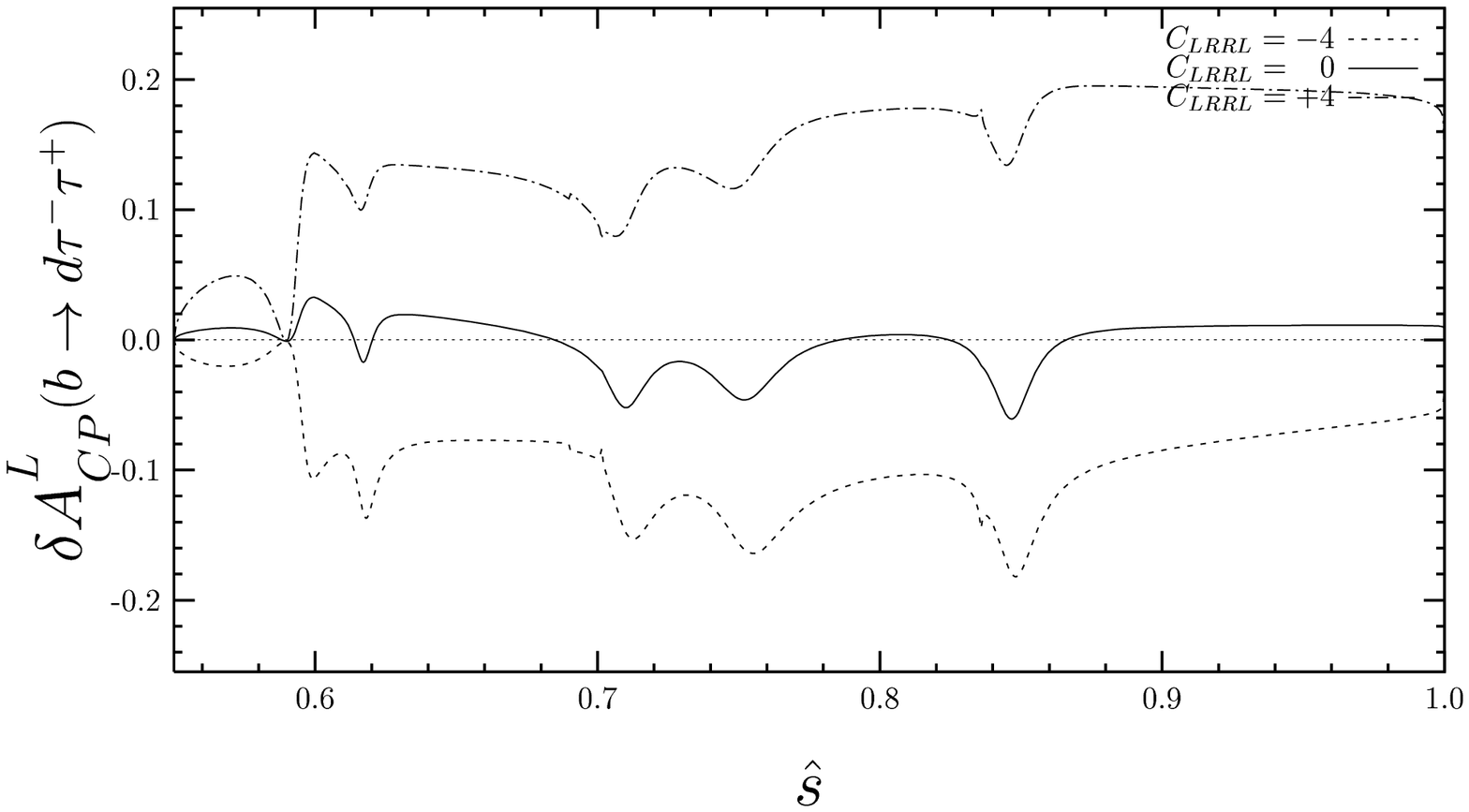}
\vskip 7.8 cm
\caption{}
\end{figure}

\begin{figure}
\vskip 1.5 cm
    \includegraphics{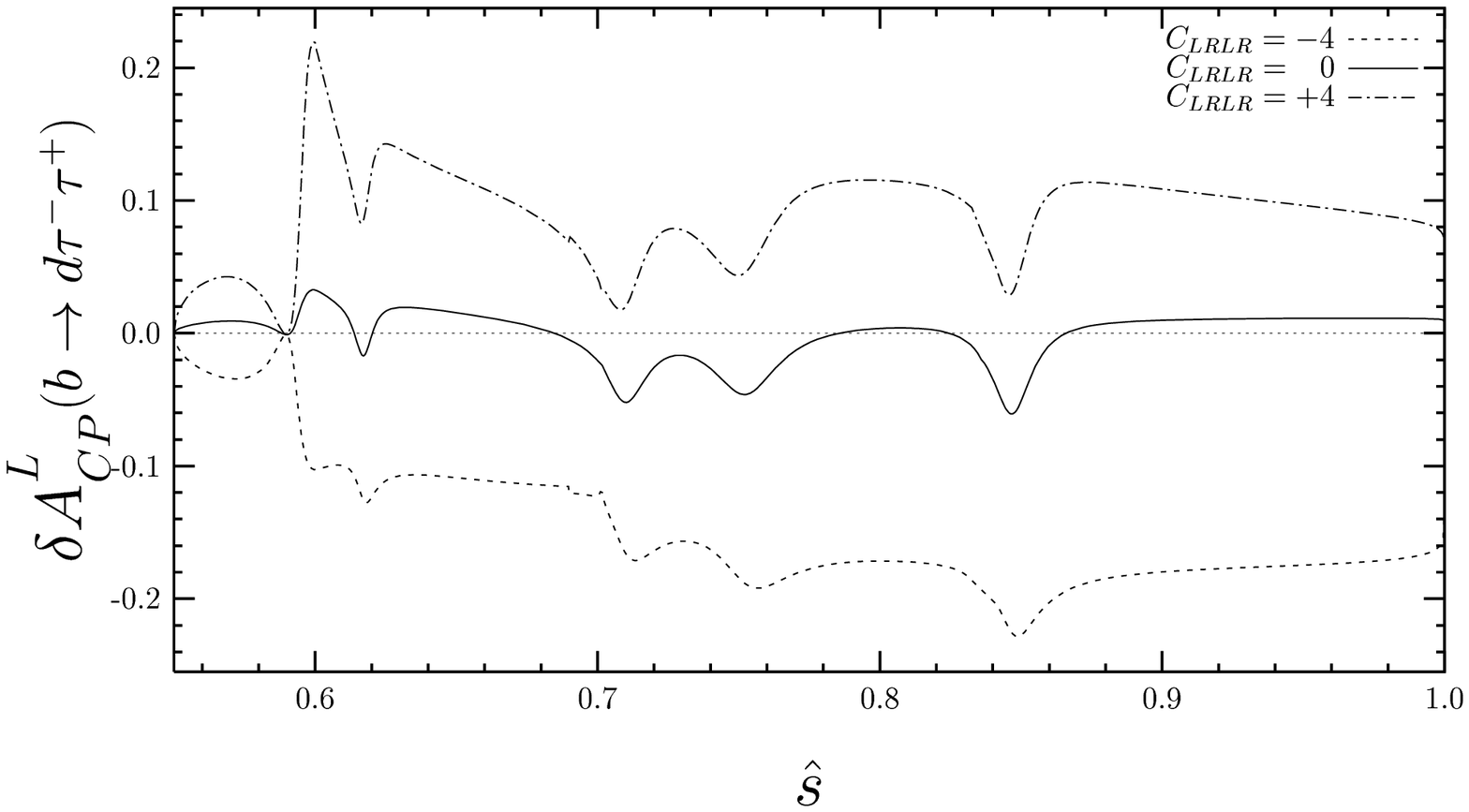}
\vskip 7.8cm
\caption{}
\end{figure}

\begin{figure}
\vskip 2.5 cm
    \includegraphics{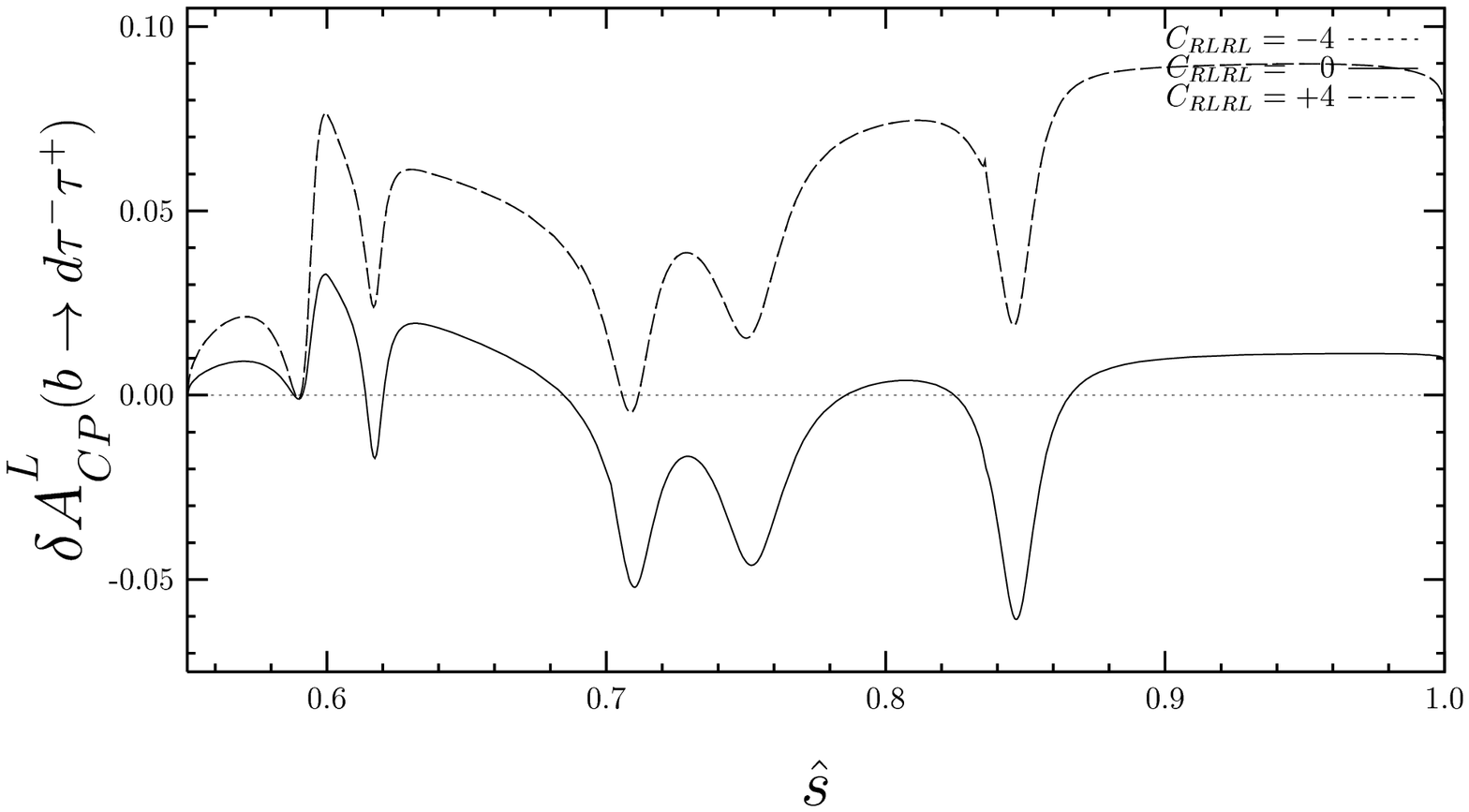}
\vskip 7.8 cm
\caption{}
\end{figure}

\begin{figure}
\vskip 1.5 cm
    \includegraphics{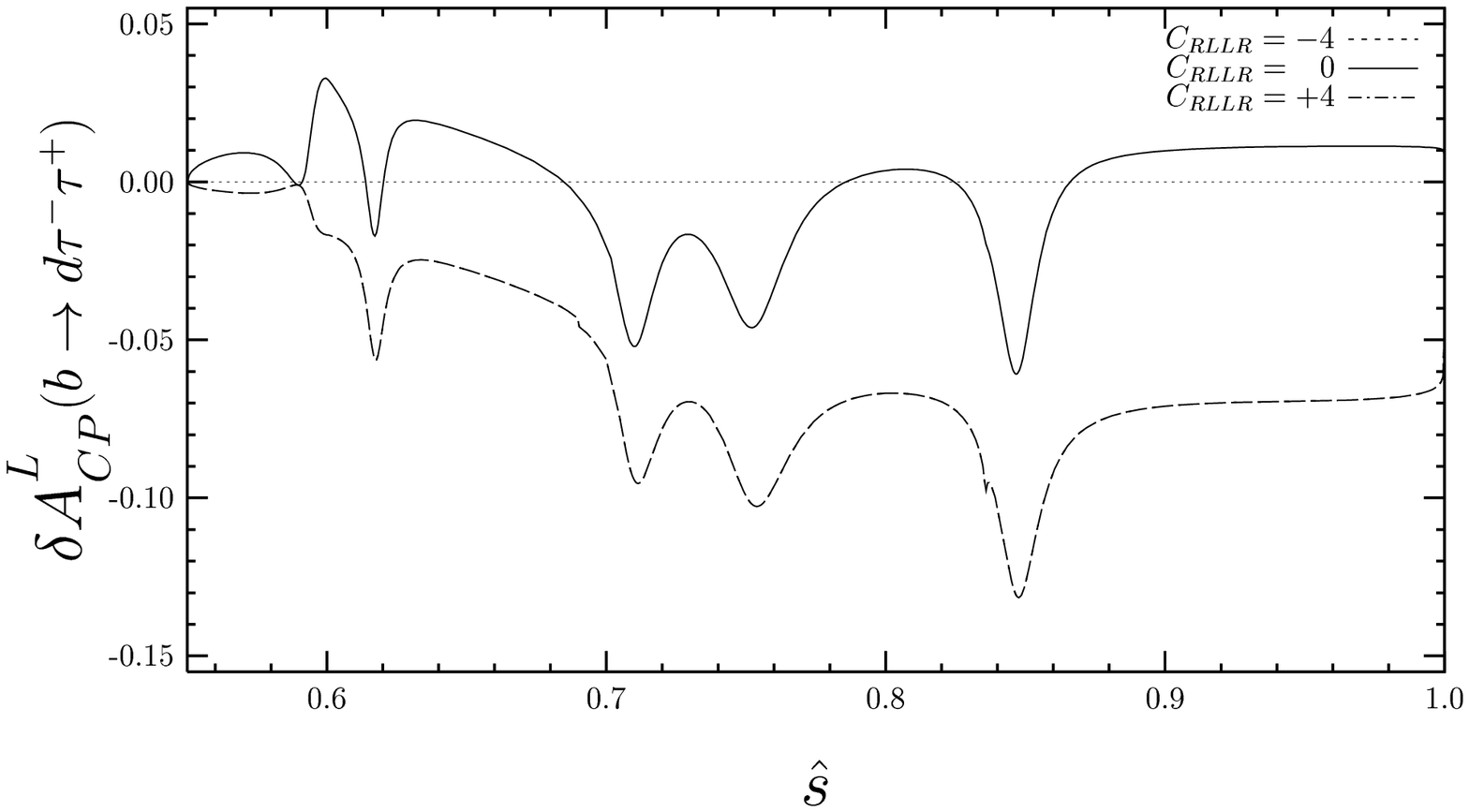}
\vskip 7.8cm
\caption{}
\end{figure}

\begin{figure}
\vskip 2.5 cm
    \includegraphics{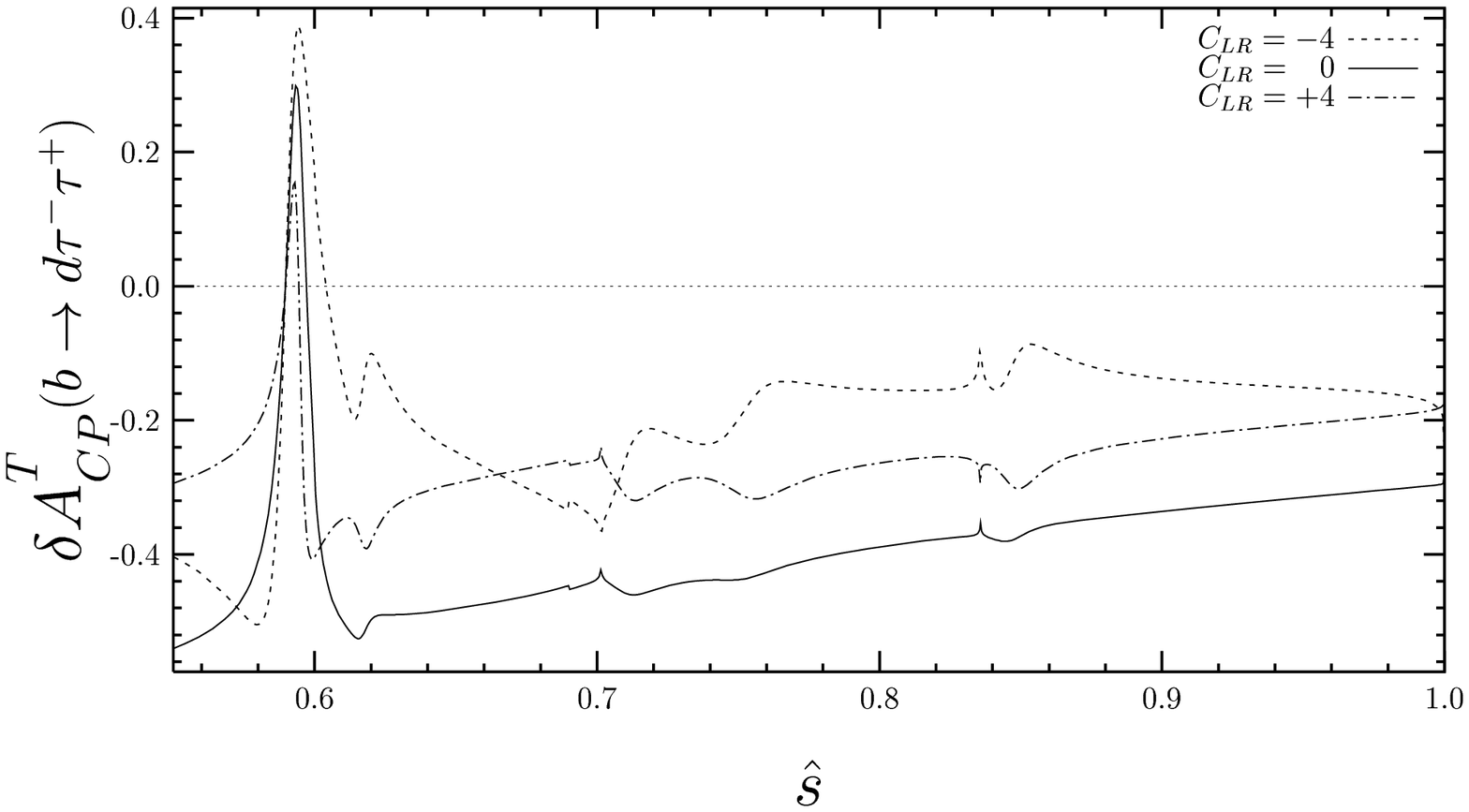}
\vskip 7.8 cm
\caption{}
\end{figure}

\begin{figure}
\vskip 1.5 cm
    \includegraphics{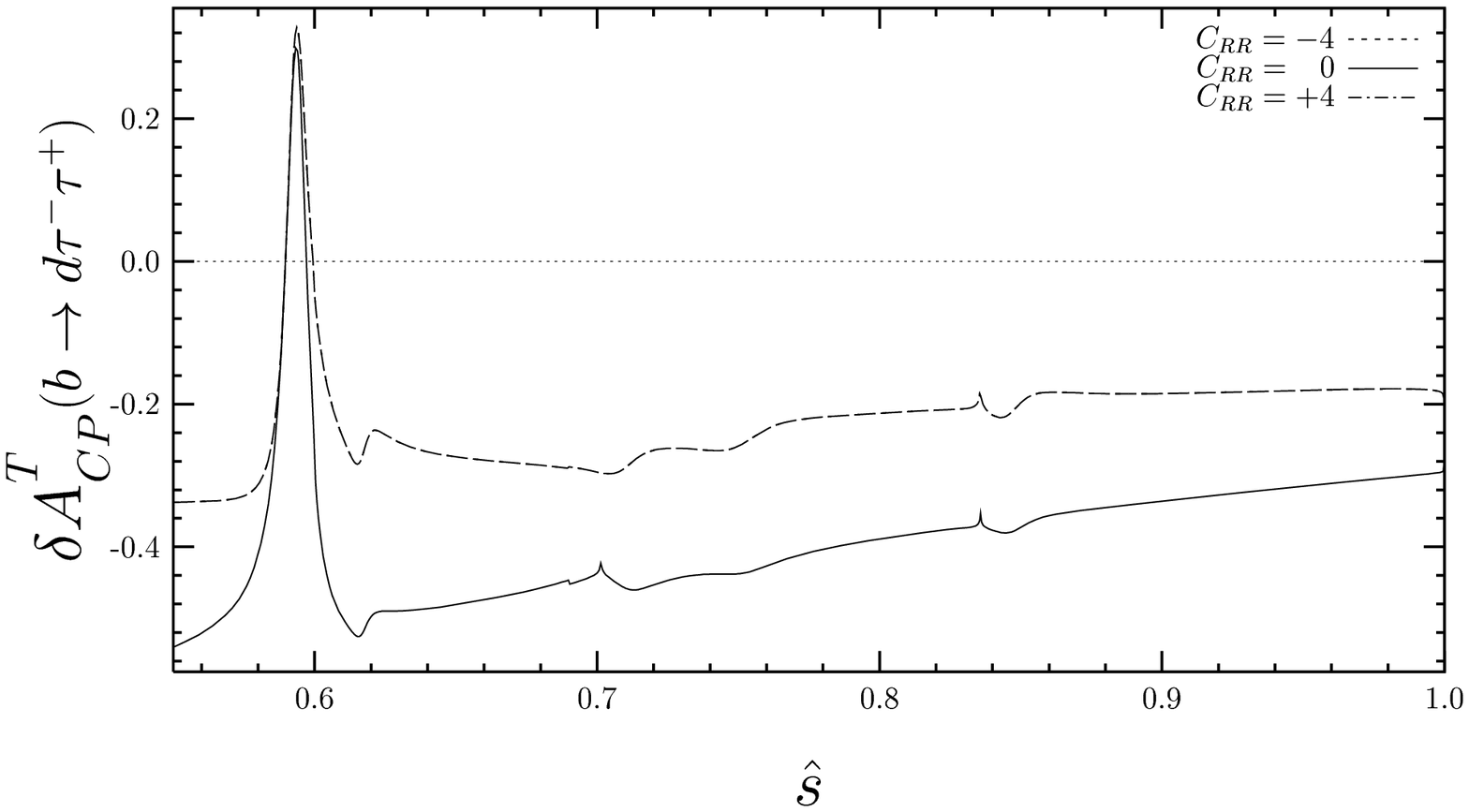}
\vskip 7.8cm
\caption{}
\end{figure}

\begin{figure}
\vskip 2.5 cm
    \includegraphics{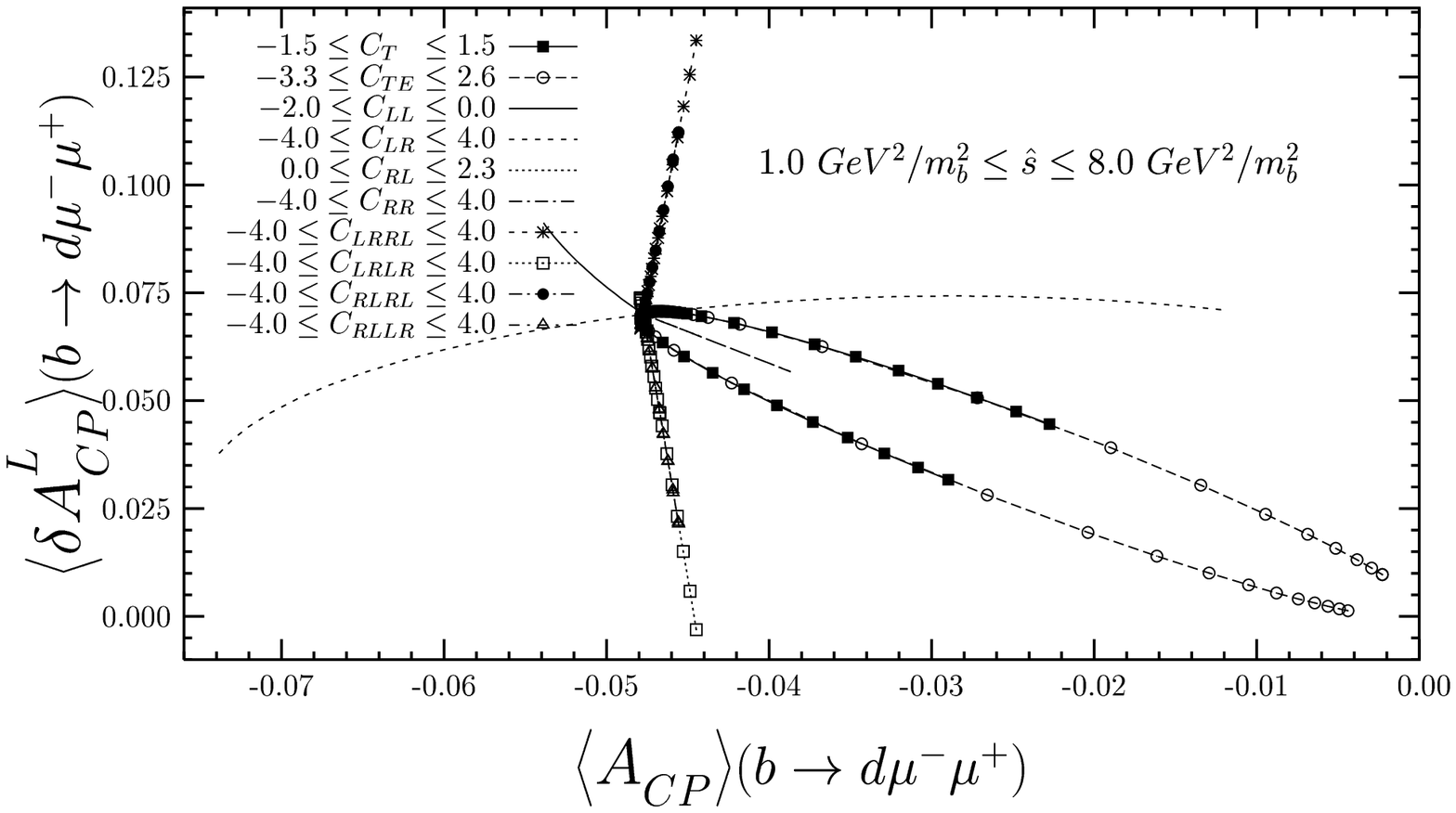}
\vskip 7.8 cm
\caption{}
\end{figure}

\begin{figure}
\vskip 1.5 cm
    \includegraphics{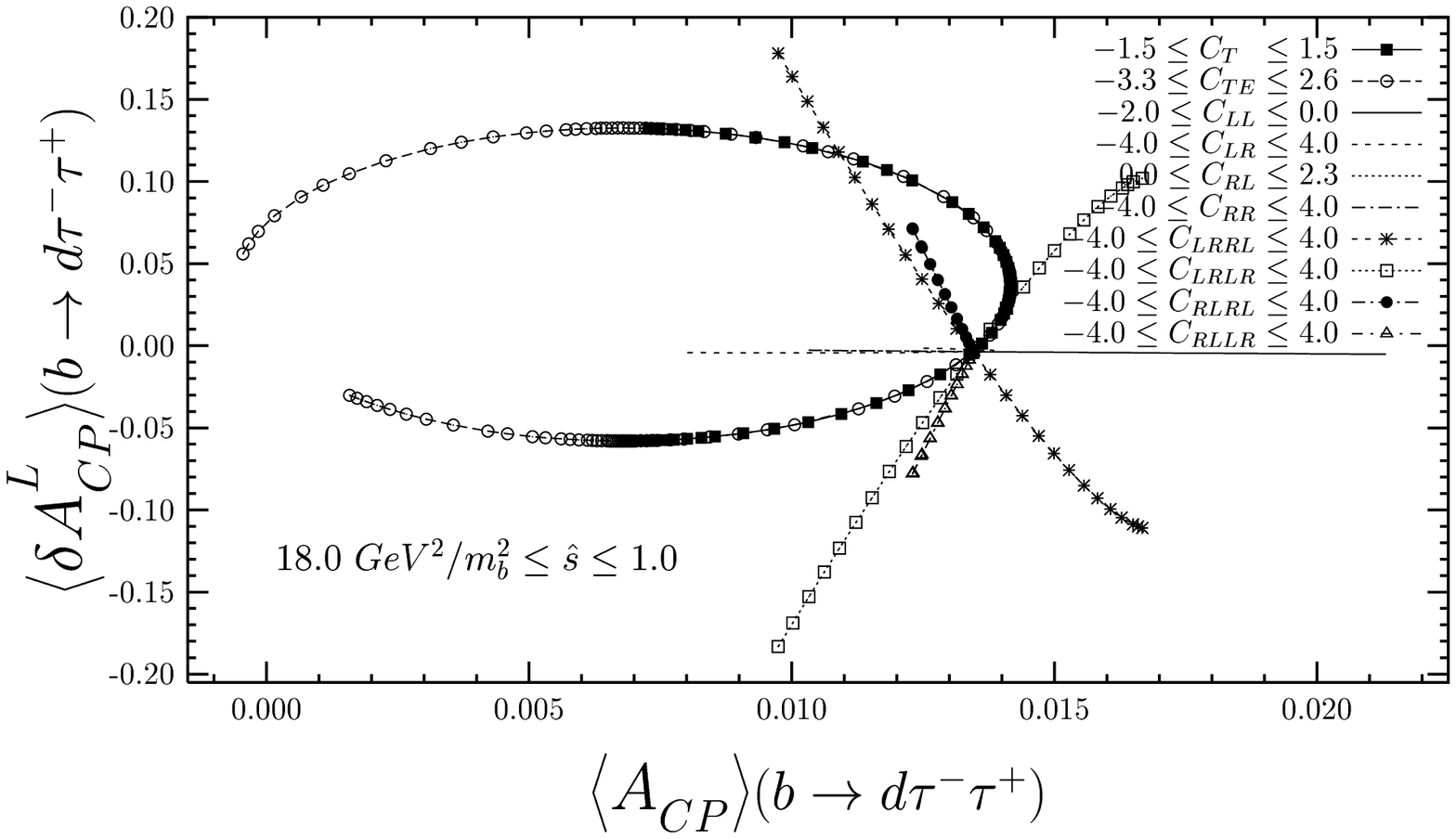}
\vskip 7.8cm
\caption{}
\end{figure}

\begin{figure}
\vskip 2.5 cm
    \includegraphics{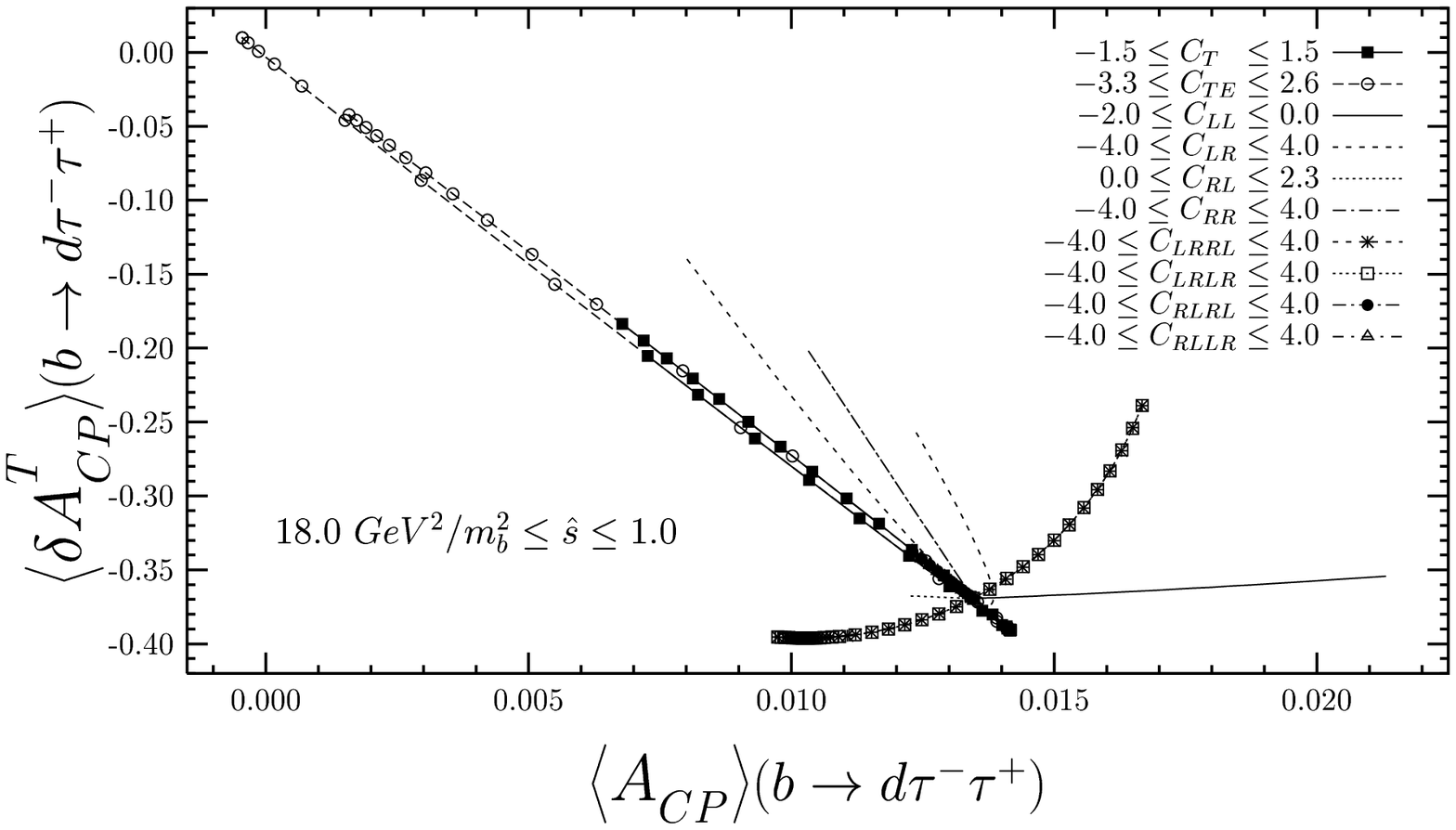}
\vskip 7.8 cm
\caption{}
\end{figure}

\end{document}